\documentclass{article}
\usepackage{graphicx}  
\usepackage{amsmath}   
\usepackage[compress]{cite}
\usepackage{amssymb}   
\usepackage{bm} 
\usepackage{dcolumn}
\usepackage{color}
\usepackage{mathrsfs}
\usepackage{gensymb}
\usepackage{float}
\usepackage{amsfonts}
\usepackage{varioref}
\usepackage{textcomp}
\usepackage[caption=true]{subfig}
\RequirePackage[colorlinks,citecolor=blue,urlcolor=magenta,linkcolor=blue]{hyperref}
\allowdisplaybreaks
\def\gr{general relativity}

\labelformat{section}{Section #1} 
\labelformat{subsection}{Section #1} 
\labelformat{subsubsection}{Section #1}
\labelformat{subsubsubsection}{Section #1}
\labelformat{equation}{Eq.~(#1)} 
\labelformat{figure}{Fig.~#1} 
\labelformat{subfigure}{Fig.~\thefigure#1} 
\labelformat{table}{Tab.~#1} 
\labelformat{appendix}{Appendix #1}
\title{Signatures of regular black holes from the shadow of Sgr A* and M87*}
\author{Indrani Banerjee\footnote{banerjeein@nitrkl.ac.in}~$^{1}$, Subhadip Sau\footnote{subhadipsau2@gmail.com}~$^{2,3}$ and Soumitra SenGupta\footnote{tpssg@iacs.res.in}~$^{3}$\\
{$^{1}$\small{Department of Physics and Astronomy, National Institute of Technology, Rourkela}}\\
{$^{2}$\small{Department of Physics, Jhargram Raj College, Jhargram, West Bengal-721507}}\\
{$^{3}$\small{School of Physical Sciences, Indian Association for the Cultivation of Science, Kolkata-700032}}}
\date{ }  
\begin{document}
  
\maketitle
\begin{abstract}
With the recent release of the black hole image of Sgr A* alongside the earlier image of M87*, one can now really hope to acquire a better understanding of the gravitational physics at the horizon scale. In this paper, we investigate the prospect of the regular black hole scenario with a Minkowski core in explaining the observed shadow of M87* and Sgr A*. Regular black holes generally appear in Einstein gravity coupled to non-linear electrodynamics and are interesting as they can evade the $r=0$ curvature singularity arising in \gr. Using the previously determined mass and distance we compute the observables associated with the black hole shadow. These when compared with the observed angular diameter reveals that the shadow of M87* and Sgr A* favor the regular black hole scenario with a small but non-zero charge. The implications are discussed.

\end{abstract}
\section{Introduction}

Discovery of gravitational waves \cite{Abbott:2016blz,TheLIGOScientific:2017qsa} and successful imaging of black holes by the Event Horizon Telescope collaboration \cite{Fish:2016jil,Akiyama:2019cqa,Akiyama:2019brx,Akiyama:2019sww,Akiyama:2019bqs,Akiyama:2019fyp,Akiyama:2019eap} has opened up new avenues for testing \gr\ (GR) in the strong field regime. This is important, since \gr\, despite being the most successful theory in explaining gravitational interaction in all length scales has several shortcomings. In the observational side GR fails to adequately address the dark sector \cite{Milgrom:1983pn,Bekenstein:1984tv,Perlmutter:1998np,Riess:1998cb} while in the theoretical end, the theory is marred with unresolved issues like singularities \cite{Penrose:1964wq,Hawking:1976ra,Christodoulou:1991yfa}. Therefore, the need to understand the nature of strong gravity continues to be a subject of considerable importance. 

Singularities are regions of spacetime where a theory loses its predictability and therefore it is believed that singularities should not exist in nature. However, the theorems proposed by Hawking and Penrose \cite{Hawking:1973uf} state that singularities are inevitable in GR.   It is often believed that a suitable theory of quantum gravity can resolve this issue and hence a number of quantum gravity models have been put forward \cite{Horava:1995qa,Horava:1996ma,Polchinski:1998rq,Polchinski:1998rr,Ashtekar:2006rx,Kothawala:2013maa}. An alternate route to address the singularity issue classically, is by studying black holes with a regular core \cite{Ayon-Beato:2000mjt,Ayon-Beato:1998hmi,Ayon-Beato:2004ywd,Bronnikov:2000vy,Borde:1996df,Barrabes:1995nk,Ayon-Beato:1999qin,Bonanno:2000ep,Nicolini:2005vd,Myung:2007qt,PhysRevLett.96.031103}. For such black holes the curvature invariants are finite for all points in space time. These black holes have a horizon and the core at $r=0$ is often of de Sitter or Minkowski type \cite{Frolov:1988vj,Mukhanov:1991zn,Brandenberger:1993ef,Ansoldi:2008jw,1966JETP...22..378G,Fan:2016hvf}.  

Regular black holes with a de Sitter core are more extensively studied \cite{Ansoldi:2008jw,1966JETP...22..378G,Fan:2016hvf}. This motivates us study the less explored regular black holes with a Minkowski core. Such black holes arise in gravity theories coupled to non-linear electrodynamics where the source of Einstein's equations correspond to an anisotropic fliud resembling Maxwell's stress tensor far from the black hole \cite{Culetu:2014lca}. As a result the metric resembles the Reissner N\"{o}rsdrom spacetime and the electric field assumes the form of the Coulomb field far from the source. Investigating the properties of such a black hole solution is important as the mass function is associated with an exponential convergence factor which makes the corresponding quantum gravity model finite to all orders \cite{Brown:1980uk}. A finite quantum gravity model is desirable as it can evade the cosmological constant problem \cite{Moffat:2001jf} and avoid the divergences appearing in flat space quantum field theories. 

Astrophysical black holes are generally rotating and since we wish to investigate the observational signatures of the aforesaid regular black hole, studying its rotating counterpart is important. Such a rotating solution is obtained by applying the Newman-Janis algorithm \cite{Newman:1965tw,Azreg-Ainou:2014pra,Azreg-Ainou:2014aqa,Azreg-Ainou:2014nra} to the static, spherically symmetric seed metric which has an exponential mass function. Just as the spherically symmetric solution resembles the Reissner N\"{o}rsdrom metric, the axisymmetric solution resembles the Kerr-Newman background far from the source \cite{Ghosh:2014pba}.

The observational signatures of regular black holes have been studied quite extensively \cite{Kumar:2020ltt,Kumar:2020yem,Allahyari:2019jqz,Vagnozzi:2022moj,Uniyal:2022vdu,Stuchlik:2014qja,Schee:2015nua,Schee:2016mjd,Stuchlik:2019uvf,Schee:2019gki,Banerjee:2022ffu,Banerjee:2021nza,Banerjee:2022chn}.
In this work we investigate the nature of the black hole shadow cast by the regular black holes discussed above. 
We derive the outline of the black hole shadow corresponding to the rotating solution which enables us to compute the observables like the angular diameter of the shadow, the deviation from circularity and the axis ratio. These are then compared with the observed image of M87* \cite{Fish:2016jil,Akiyama:2019cqa,Akiyama:2019brx,Akiyama:2019sww,Akiyama:2019bqs,Akiyama:2019fyp,Akiyama:2019eap} and Sgr A* \cite{EventHorizonTelescope:2022xnr,EventHorizonTelescope:2022vjs,EventHorizonTelescope:2022wok,EventHorizonTelescope:2022exc,EventHorizonTelescope:2022urf,EventHorizonTelescope:2022xqj} which in turn enables us to establish novel constrains on the non-linear electrodynamics (NED) charge parameter of the two sources. Estimate of the NED charge parameter for some stellar mass black holes and Sgr A* has been done from observations related to quasi-periodic oscillations. The present analysis therefore opens up the opportunity to verify their mutual consistency. 

The paper is organized in the following way: In \ref{Sec2} we briefly discuss the characteristics of the regular black hole solution with a Minkowski core arising in gravity coupled to non-linear electrodynamics. \ref{Sec3} is dedicated in deriving the outline of the black hole shadow for the regular spacetime studied in \ref{Sec2}. In \ref{Sec4} we compare our theoretical results with the observed shadow of M87* and Sgr A* respectively, in particular, we compute observables like the shadow angular diameter, the deviation from circularity and the axis ratio which in turn enables us to conclude whether the regular black hole scenario is more favored compared to the Kerr scenario in GR. Finally we conclude with a summary of our results with some scope for future work in \ref{Sec6}.

Here we work with mostly positive metric convention and assume $G=c=1$ for our analysis.

\section{Black hole in non-linear electrodynamics}\label{Sec2}
The present work deals with regular black holes in non-linear electrodynamics with an asymptotically Minkowski core. The Lagrangian density associated with non-linear electrodynamics coupled to Einstein gravity is given by \cite{Kumar:2020ltt,Ayon-Beato:2000mjt,Salazar:1987ap,Fan:2016hvf,Bronnikov:2017tnz,Toshmatov:2018cks}, 
\begin{align}
S=\int d^4 x \sqrt{-g} \bigg(\frac{\mathcal{R}}{16\pi}-\frac{L(F)}{4\pi} \bigg)
\label{Eq1}
\end{align}
where $\mathcal{R}$ denotes the Ricci scalar and $L(F)$ is the non-linear electrodynamics Lagrangian density. In \ref{Eq1} $F=F^{ab}F_{ab}/4$ corresponds to the Faraday invariant while $F_{ab}=\partial_a A_b - \partial_b A_a$ is the electromagnetic field strength tensor with $A_i$ the gauge field. In the weak field limit $L(F)=F$ and the Maxwell theory is retrieved. By varying the action with respect to $A_i$ we obtain its corresponding equation of motion,
\begin{align}
\lbrace L_F F^{ij}\rbrace_{;i}=0 ~~~~(^*F^{ij})_{;i}=0
\label{Eq3}
\end{align} 
where $L_F=\frac{\partial L }{\partial F}$ and $^*F^{ij}=\epsilon^{ijkl}F_{kl}$ is the Hodge-dual of $F^{ab}$. On the other hand, variation of the action with respect to the metric leads to the Einstein's equations with $L(F)$ as the source,
\begin{align}
G_{ab}=2 (L_F F_a^s F_{bs}-g_{ab}L(F))
\label{Eq2}
\end{align}
The first regular black hole solution was proposed by Bardeen \cite{Bardeen:1968cux} which eventually led to extensive work in this direction \cite{Ayon-Beato:1998hmi,Ayon-Beato:1999qin,Dymnikova:1992ux,Dymnikova:2004zc,Bronnikov:2000vy,Bronnikov:2005gm,Burinskii:2002pz,Berej:2006cc,Balart:2014cga,Sajadi:2017glu,Ghosh:2018bxg}. Later Ayon-Beato and Garcia confirmed that the physical source associated with the Bardeen solution corresponds
to the gravitational field of a nonlinear magnetic monopole of a self-gravitating magnetic field \cite{Ayon-Beato:2000mjt}. In general, the static and spherically symmetric solution of \ref{Eq2} assumes the form \cite{Kumar:2020ltt},
\begin{align}
ds^2=-\Big(1-\frac{2\hat{m}(r)}{r}\Big)dt^2 + \Big(1-\frac{2\hat{m}(r)}{r}\Big)^{-1}dr^2 + r^2 d\theta^2 + r^2 sin^2\theta d\phi^2
\label{Eq-4}
\end{align}
where the form of the mass function $\hat{m}(r)$ is determined by the properties of the non-linear electrodynamics source $L(F)$. For a spherically symmetric spacetime the non-zero components of the field strength tensor $F_{ab}$ correspond to $F_{tr}$ and $F_{\theta\phi}$. In the event the black hole carries a pure magnetic charge $g$ the gauge field potential corresponds to $A=-gcos\theta d\phi$ such that the only non-vanishing component of the field strength tensor is $F_{\theta\phi}=gsin\theta$.

The Lagrangian density associated with non-linear electrodynamics leading to regular black holes with a monopole charge has the general form \cite{Kumar:2020ltt},
\begin{align}
L(F)=\frac{\mu \mathcal{M}}{g^3}\frac{(2g^2 F)^{\frac{\nu+3}{4}}}{\lbrace1+(2g^2F)^\frac{\nu}{4}\rbrace^{1+\frac{\mu}{\nu}}}
\label{5}
\end{align}
Using the above Lagrangian density in \ref{Eq2}, we obtain the following form of the mass function,
\begin{align}
\hat{m}(r)=\frac{\mathcal{M} r^\mu}{(r^\nu + g^\nu)^{\mu/\nu}}
\label{Eq-6}
\end{align}
where $\mathcal{M}$ is the black hole mass while $\mu$ and $\nu$ are positive, dimensionless constants appropriately chosen to ensure asymptotic flatness. It is important to note that the Schwarzschild solution is regained when $\mu$ vanishes. The mass function in \ref{Eq-6} gives rise to a repulsive de Sitter central core   
for $\mu \geq 3$ and these two constants ($\mu$ and $\nu$) can be suitably chosen to give rise to exact spherically symmetric regular black hole solutions. For example, $\mu=3$ and $\nu=2$ gives rise to the well-known Bardeen black hole solution while $\mu=\nu=3$ leads to the Hayward black hole solution.

The black holes discussed so far are endowed with an asymptotically de Sitter core whereas in this work we are interested in regular black holes with an asymptotically Minkowski core.
Such black holes have been proposed by Ayon-Beato and Garcia with mass function \cite{Ayon-Beato:1998hmi},
\begin{align}
\hat{m}(r)=\frac{\mathcal{M}r^3}{(r^2 + q^2)^\frac{3}{2}}-\frac{q^2r^3}{2(r^2 + q^2)^2}
\label{7}
\end{align}
and Lagrangian density,
\begin{align}
L(F)=P\frac{(1-8\sqrt{-2q^2P}-6q^2P)}{(1+\sqrt{-2q^2P})^4}-\frac{3}{4q^2s}\frac{(-2q^2P)^\frac{5}{4}(3-2\sqrt{-2q^2P})}{(1+\sqrt{-2q^2P})^\frac{7}{2}}
\label{8}
\end{align} 
where $q$ is the electric charge of the black hole, $s=\frac{|q|}{2\mathcal{M}}$ and $P=L_F^2F$. 
The same authors also proposed another class of regular black hole with an asymptotically Minkowski core \cite{Ayon-Beato:1999kuh}. Such black holes are associated with the mass function,
\begin{align}
\hat{m}(r)=\mathcal{M}\bigg\lbrace 1 -tanh\bigg(\frac{q^2}{2\mathcal{M}r}\bigg)\bigg\rbrace
\label{9}
\end{align}
which is obtained by solving \ref{Eq2} with	the Lagrangian density,
\begin{align}
L=2PH_P-H
\label{10}
\end{align}
where, 
\begin{align}
H=P\lbrace 1- tanh^2(s\sqrt[4]{-2q^2P})\rbrace
\end{align}
and $H_P=\frac{\partial H}{\partial P}$. 

In the present work the Lagrangian density corresponding to non-linear electrodynamics is given by \cite{Kumar:2020ltt},
\begin{align}
L(F)=Fe^{-\alpha(2g^2F)^{1/4}}
\label{Eq4}
\end{align}
such that $\alpha=g/(2\mathcal{M})$ with $g$ being the magnetic charge and $\mathcal{M}$ the mass of the black hole. 

With the Lagrangian density in \ref{Eq4}, the static, spherically symmetric and asymptotically flat solution of \ref{Eq2} assumes the form,
\begin{align}
ds^2=-\Bigg(1-\frac{2 \mathcal{M}e^{-\kappa/r}}{r}\Bigg)c^2dt^2 + \frac{dr^2}{\Bigg(1-\frac{2\mathcal{M}e^{-\kappa/r}}{r}\Bigg)} +r^2 (d\theta^2 + sin^2\theta d\phi^2)
\label{Eq5}
\end{align}
where $\kappa=g^2/2\mathcal{M}$.

Such a black hole solution arises as a result of solving Einstein's equations with the source \cite{Culetu:2013fsa},
\begin{align}
&T^0_0=-\rho(r)=\frac{-\mathcal{M} k}{4\pi  r^4} e^{-k/{r}};\nonumber \\
&T^1_1=-\rho(r)=\frac{-\mathcal{M} k}{4\pi  r^4} e^{-k/{r}}; \nonumber \\
&T^2_2=T^3_3=\frac{\mathcal{M} k}{4\pi  r^4}\Bigg(1-\frac{k}{2r}\Bigg)e^{-k/r}
\label{energy}
\end{align}
where $k=\kappa/r_g$ is the dimensionless magnetic monopole charge parameter with $r_g=GM/c^2$ the gravitational radius. 
It is interesting to note that the above energy momentum tensor is regular at $r=0$ and vanishes as $r\to \infty$. Moreover, \ref{energy} is in accordance with the weak energy condition and reduces to the Maxwell stress tensor far from the horizon.

Since astrophysical black holes are in general rotating, studying the axisymmetric counterpart of \ref{Eq5} is observationally more relevant. The stationary, axisymmetric and asymptotically flat black hole solution of Einstein's equations with source given by \ref{energy} is obtained by applying the Newman-Janis algorithm \cite{Newman:1965tw,Azreg-Ainou:2014pra,Azreg-Ainou:2014aqa,Azreg-Ainou:2014nra} to the seed metric \ref{Eq5} \cite{Ghosh:2014pba}. The corresponding line element has been studied in \cite{Ghosh:2014pba,Kumar:2020ltt} and assumes the form,
\begin{align}
\label{metric_bardeen}
ds^{2} &=-\bigg{(} 1 - \frac{2\hat{m}(r)r}{{\Sigma}}\bigg{)}dt^{2} - \frac{4a\hat{m}(r)r}{{\Sigma}}\sin^{2}\theta dt d\phi + \frac{{\Sigma}}{\Delta}dr^{2} \nonumber\\
&+{\Sigma} d\theta^{2} + \bigg{(} r^{2} + a^{2} + \frac{2\hat{m}(r)ra^{2}}{{\Sigma}}\sin^{2}\theta\bigg{)}\sin^{2}\theta d\phi^{2}
\end{align}
where, 
\begin{align}\label{metricparams_bardeen}
{\Sigma} = r^{2} + a^{2}\cos^{2}\theta ~ {,} ~ \Delta = r^{2} + a^{2} - 2\hat{m}(r)r 
\end{align}
and $a$ is the Kerr parameter.
The mass function $\hat{m}(r)$ is given by,
\begin{align}
\hat{m}(r)=\mathcal{M}e^{-\kappa/r}
\label{Eq-1}
\end{align}
such that $lim_{r\rightarrow\infty}\hat{m}(r) ={\mathcal{M}}$. 
In \ref{Eq-1} $\kappa=\frac{g^2}{2\mathcal{M}}$ has dimensions of length. Since it is computationally easier to handle dimensionless quantities we scale $\kappa$ and $r$ in \ref{Eq-1} by the gravitational radius $r_g=GM/c^2$. Thus the dimensionless metric parameters correspond to the squared charge to mass ratio $k=\kappa/r_g=\frac{g^2c^4}{2G^2\mathcal{M}^2}$ and the spin parameter $a\equiv a/r_g$.

It is important to note that when $r>>k$ the above metric reduces to the Kerr-Newman spacetime. In the absence of nonlinear electrodynamics $k=0$ and \ref{metric_bardeen} reduces to the Kerr metric. We further note that \ref{metric_bardeen} has no curvature singularity at $r=0$ but assumes an asymptotically Minkowski core, i.e. the energy density $\rho(r)\to 0$ as $r\to 0$ unlike a de-Sitter core where the energy density becomes constant at the core. Interestingly, the curvature invariants in the above spacetime can be described by the Lambert W function and exhibits several physically intriguing features \cite{Valluri:2000zz,Boonserm:2008zg,Boonserm:2010px,Boonserm:2013dua,Boonserm:2018orb,Sonoda:2013kia,Sonoda:2013jia,Culetu:2013fsa}.

In order to derive the horizon radii one solves for the roots of $g^{rr}=\Delta=0$ which yields,
\begin{align}
r^2 + a^2 - 2re^{-k/r} =0
\label{horizon_bardeen}
\end{align}
For the metric in \ref{metric_bardeen} to represent a black hole the horizons must be real and positive. This requirement sets the physically allowed range of $k$ to $0\lesssim k \lesssim 0.7$.
In the next section we discuss the procedure to derive the outline of the shadow for the regular spacetime given in \ref{metric_bardeen}.

\section{Shadow of regular black holes with Minkowski core}\label{Sec3}
The boundary of the shadow bears the signatures of strong gravitational lensing of nearby radiation and therefore the nature of the shadow can potentially unravel valuable information regarding the characteristics of strong gravity near the black hole \cite{Gralla:2019xty,Bambi:2019tjh,Hioki:2009na,Vagnozzi:2019apd,Banerjee:2019nnj}.
In this section we derive the nature of the black hole shadow for black holes discussed in \ref{Sec2} \cite{Cunha:2018acu,Vries_1999}. Given a stationary, axisymmetric metric, the Lagrangian $\tilde{\mathcal{L}}$ for test particle motion is given by,
\begin{align}
g_{\mu\nu} \dot{x}^{\mu} \dot{x}^{\nu}=g_{tt}\dot{t}^2 + 2g_{t\phi}\dot{t}\dot{\phi}+g_{\phi\phi}\dot{\phi}^2 + g_{rr} \dot{r}^2 + g_{\theta\theta}\dot{\theta}^2 =2\tilde{\mathcal{L}}
\label{1}
\end{align}
The Lagrangian is equal to unity for massive particles and zero for massless particles. The corresponding Hamiltonian is given by,
\begin{align}
\mathcal{H}=p_\mu \dot{x^\mu}- \tilde{\mathcal{L}}=\frac{1}{2} g^{\mu\nu}p_\mu p_\nu=\frac{k}{2}
\label{2}
\end{align}
with $k$ the rest mass of the test particle which in the present context is zero. We use the Hamilton-Jacobi approach such that the Hamiltonian is related to the action $S$ by,
\begin{align}
\mathcal{H}(x^\mu, p^\mu)+\frac{\partial S}{\partial \lambda}=0~~~~\rm where~~~~~~~~~~~~~~~~~p_\mu=\frac{\partial S}{\partial x^\mu}
\label{3}
\end{align}
Since the metric in \ref{1} is independent of $t$ and $\phi$, the specific energy $\mathcal{E}$ and specific angular momentum $\mathcal{L}$ are conserved quantities. These are given by,
\begin{align}
\mathcal{E}=-p_t=g_{tt}\dot{t}+g_{t\phi}\dot{\phi}=\rm constant\nonumber\\ 
&\mathcal{L}=p_\phi=g_{\phi t}\dot{t} + g_{\phi\phi}\dot{\phi}=\rm constant
\label{4}
\end{align}
From the above condition the action can be written as, 
 \begin{align}
S= -\mathcal{E}t + \mathcal{L}\phi + S(r,\theta)
\label{5}
\end{align}
It turns out that for the metric in \ref{metric_bardeen}, \ref{5} is separable such that, $S(r,\theta)=S^r(r) + S^\theta(\theta)$. Substituting \ref{5} in equation \ref{2} we get,
\begin{align}
  g^{rr}\bigg(\frac{d S}{d r}\bigg)^2 + g^{\theta\theta}\bigg(\frac{d S}{d \theta}\bigg)^2 + g^{tt}\mathcal{E}^2-2^{t\phi}\mathcal{EL} + g^{\phi\phi}\mathcal{L}^2=0
\label{6}
\end{align}
For the metric in \ref{metric_bardeen} the above equation assumes the form, 
\begin{align}
\Delta\Bigg(\frac{dS^r}{dr}\Bigg)^2+\Bigg(\frac{dS^\theta}{d\theta}\Bigg)^2-\Bigg\lbrace	
\frac{1}{\Delta}(r^2+a^2)^2-a^2sin^2\theta \Bigg\rbrace \mathcal{E}^2 + 
\frac{4ar\hat{m}(r)}{\Delta}\mathcal{EL} + \mathcal{L}^2\Bigg(\frac{1}{sin^2\theta}-\frac{a^2}{\Delta}\Bigg)=0
\label{7}
\end{align}
Interestingly, the $r$ and $\theta$ parts of the above equation can be separated such that,
\begin{align}
\Delta\Bigg(\frac{dS^r}{dr}\Bigg)^2-\frac{1}{\Delta}(r^2+a^2)^2\mathcal{E}^2 + \frac{4ar\hat{m}(r)}{\Delta}\mathcal{EL} -\frac{a^2}{\Delta}\mathcal{L}^2=-\Bigg(\frac{dS^\theta}{d\theta}\Bigg)^2 -a^2\mathcal{E}^2sin^2\theta-\frac{\mathcal{L}^2}{sin^2\theta}=C
\label{7-1}
\end{align}
where $C$ corresponds to the Carter constant. The left hand of \ref{7-1} depends only on $r$ while the right hand side is a function of $\theta$ alone. The radial part of \ref{7-1} can be written as,
\begin{align}
\Bigg(\frac{dS^r}{dr}\Bigg)^2=\frac{R(r)}{\Delta^2}
\label{8}
\end{align}
where 
\begin{align}
R(r)=\Delta[-C -(\mathcal{L}-a\mathcal{E})^2]-\lbrace(r^2+a^2)\mathcal{E} -a\mathcal{L}\rbrace ^2
\label{8-1}
\end{align}
while the angular part can be written as,
\begin{align}
\Bigg(\frac{dS^\theta}{d\theta}\Bigg)^2=\Theta(\theta)
\label{9}
\end{align}
where,
\begin{align}
\Theta(\theta)=C+cos^2\theta\Bigg(a^2\mathcal{E}^2-\frac{\mathcal{L}^2}{sin^2\theta}\Bigg)
\label{10}
\end{align}
The action therefore assumes the form,
\begin{align}
\label{11}
S=-\mathcal{E}t + \mathcal{L}\phi + \int \frac{\sqrt{R(r)}}{\Delta}dr + \int \sqrt{\Theta(\theta)}d\theta
\end{align}
From \ref{3}, \ref{8} and \ref{9} we obtain the equations of motion for $r$ and $\theta$,
\begin{align}
\dot{r}=\frac{\sqrt{R(r)}}{\Sigma} ~~~~~~~~~~~\rm{and}~~~~~~~~~~~~~~\dot{\theta}=\frac{\sqrt{\Theta(\theta)}}{\Sigma}
\label{12}
\end{align}

Estimates of the shape and size of the shadow are associated with the geodesic motion of photons in the black hole background.
Since photons are charge neutral we have considered accretion of neutral test particles in the preceding discussion.
It is however important to note that accretion of charged particles will have an impact on observables associated with black holes. In particular, when we are studying models to describe the continuum spectrum or quasiperiodic oscillations we are dealing with accretion of ions and electrons which are charged massive particles. In the event the black hole is also charged (as in the present case) the equations of motion get substantially modified due to the interaction of the charged particles with the electromagnetic field of the black hole \cite{Stuchlik:2020rls,Zhang_2022}. 
This leads to a modification in the theoretical spectrum from the accretion disk and the epicyclic frequencies associated with the quasi-periodic oscillations \cite{Stuchlik:2020rls,Zhang_2022}. 
Since photons are uncharged they do not directly interact with the electromagnetic field of the black hole. However, the charge of the black hole modifies the background metric compared to the Kerr scenario which in turn affects the motion of the photons. In other words the electromagnetic field of the black hole affects the motion of the photons through a change in the background spacetime only.

 In order to proceed further we define two impact parameters,
\begin{align}
\label{13}
\chi=\frac{C}{E^2} ~~~~~~~~~~~~~~~~\rm{and}~~~~~~~~~~~~~~~~~~~~~ \eta=\frac{L}{E}
\end{align}
From \ref{12} one can show that the physically allowed region for the photon can never reach upto $\theta=0$. The maximum allowed value of $\theta$ denoted by $\theta_{max}$ is given by,
\begin{align}
\cos^2\theta_{max}=\frac{-(\chi+\eta^2-a^2)\pm \sqrt{(\chi+\eta^2-a^2)^2+4a^2\chi})}{2a^2}
\label{14}
\end{align}
where $\chi$ can be positive, negative or zero.

\subsection{Solving the radial equation}
The differential equation associated with the radial part is given by,
\begin{align}
\Bigg(\frac{\Sigma}{E}\Bigg)^2\dot{r}^2=\Delta\Bigg[-\chi-\Bigg(\frac{L}{E}-a\Bigg)^2\Bigg]+ (r^2+a^2-a\eta)^2=V(r)
\label{19}
\end{align}
For spherical photon orbits one needs to satisfy the condition: $V(r)=V'(r)=0$. The first condition gives:
\begin{align}
\Delta(\chi+\eta^2+a^2-2a\eta)=(r^2+a^2-a\eta)^2
\label{20}
\end{align}
while the second condition leads to,
\begin{align}
\chi +\eta^2 +a^2 -2a\eta=\frac{2}{1-\frac{\hat{m}(r)}{r}-\hat{m}'(r)}(r^2+a^2-a\eta)
\label{21}
\end{align}
Solving \ref{20} and \ref{21} we get two sets of solutions for $\eta$ and $\chi$ parametrized in terms of $r$ \cite{2003GReGr..35.1909T,Vries_1999,Cunha:2018acu},

\begin{enumerate}
\item 
\begin{flalign}
\chi &= -\dfrac{r^{4}}{a^{2}}\\
\eta &= \frac{a^2 + r^{2}}{a }
\label{22-1}
\end{flalign}
\item  
\begin{flalign}
\chi&=-\frac{r^3 \left(4 a^2 r \hat{m}'(r)-4 a^2 \hat{m}(r)+r^3 \hat{m}'(r)^2+2 r^3 \hat{m}'(r)-6 r^2 \hat{m}(r) \hat{m}'(r)-6 r^2 \hat{m}(r)+9 r \hat{m}(r)^2+r^3\right)}{a^2 \left(r \hat{m}'(r)+\hat{m}(r)-r\right)^2}\\
\eta&= \frac{a^2 r \hat{m}'(r)+a^2 \hat{m}(r)+a^2 r+r^3 \hat{m}'(r)-3 r^2 \hat{m}(r)+r^3}{a \left(r \hat{m}'(r)+\hat{m}(r)-r\right)}
\label{23-1}
\end{flalign}
\end{enumerate}

The first solution is not allowed physically \cite{2003GReGr..35.1909T,Cunha:2018acu,Vries_1999} since this leads to $\Theta(\theta)<0$ which is forbidden (see \ref{12}).
For the second solution $\chi$ may take either sign depending upon the value of $r$ and accordingly the appropriate conditions need to be satisfied.

\subsection{Deriving the shape of the black hole shadow}
\label{S3.3}
Deriving the impact parameters is important as they can be used to calculate the celestial coordinates $x=\alpha$ and $y=\beta$ of the black hole shadow as seen a distant observer at position $(r_0,\theta_0)$. The apparent perpendicular distance of the image from the axis of symmetry is associated with the $x$ coordinate while the apparent perpendicular distance
of the image from the equatorial plane is associated with the $y$ coordinate of the shadow.

\begin{figure}[h!]%
    \centering
    \hspace{-1.5cm}
    
    \subfloat[Variation of BH shadow with metric parameter $k$. Here the inclination angle is taken to be $\theta=60^{\circ}$ and the spin is assumed to be $a=0.2$.]{{\includegraphics[width=7.5cm]{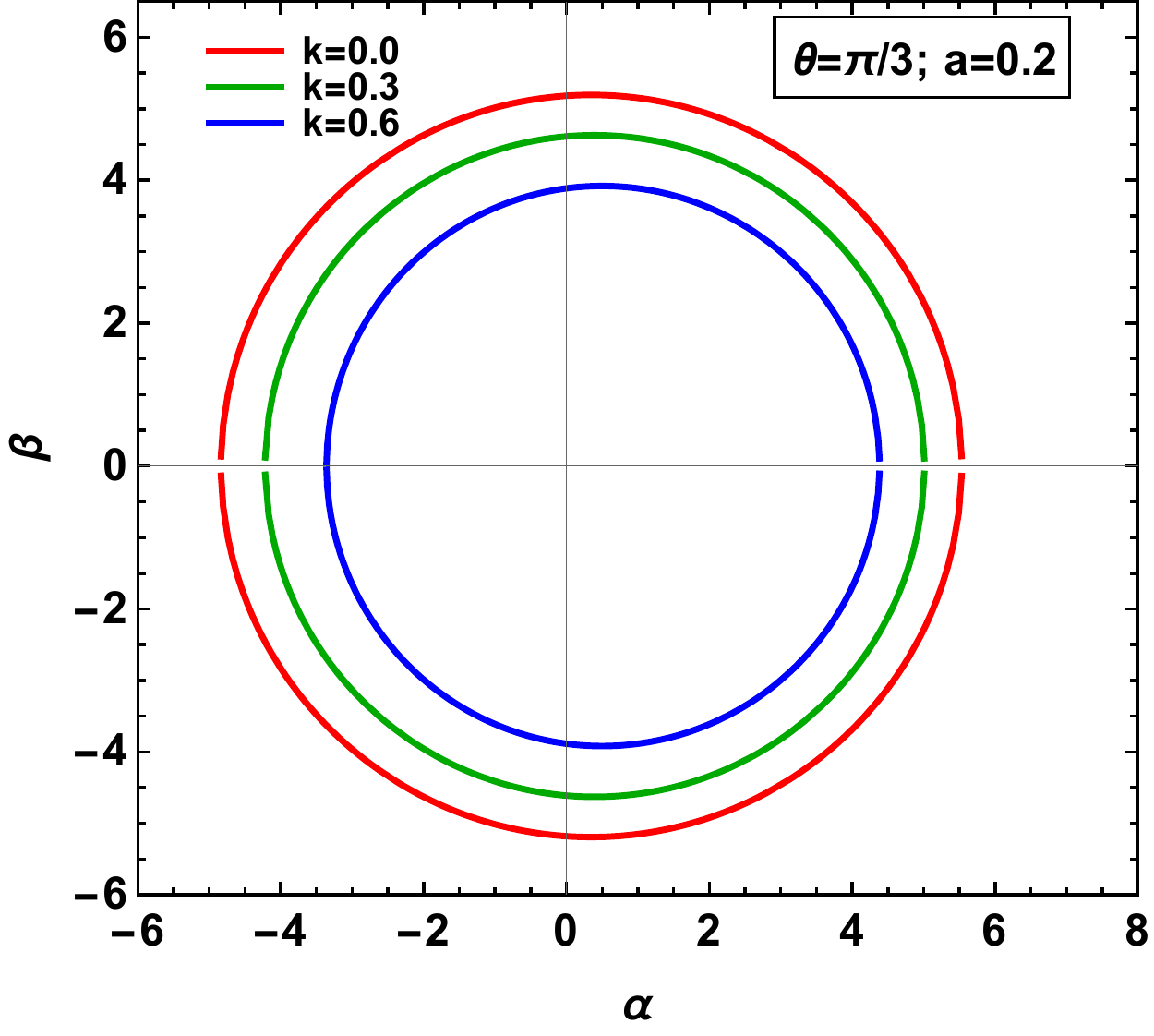} } \label{SP3_Fig_V2}}
  \qquad
     \subfloat[Variation of BH shadow with metric parameter $k$. Here the inclination angle is taken to be $\theta=45^{\circ}$ and the spin is assumed to be $a=0.5$. ]{{\includegraphics[width=7.5cm]{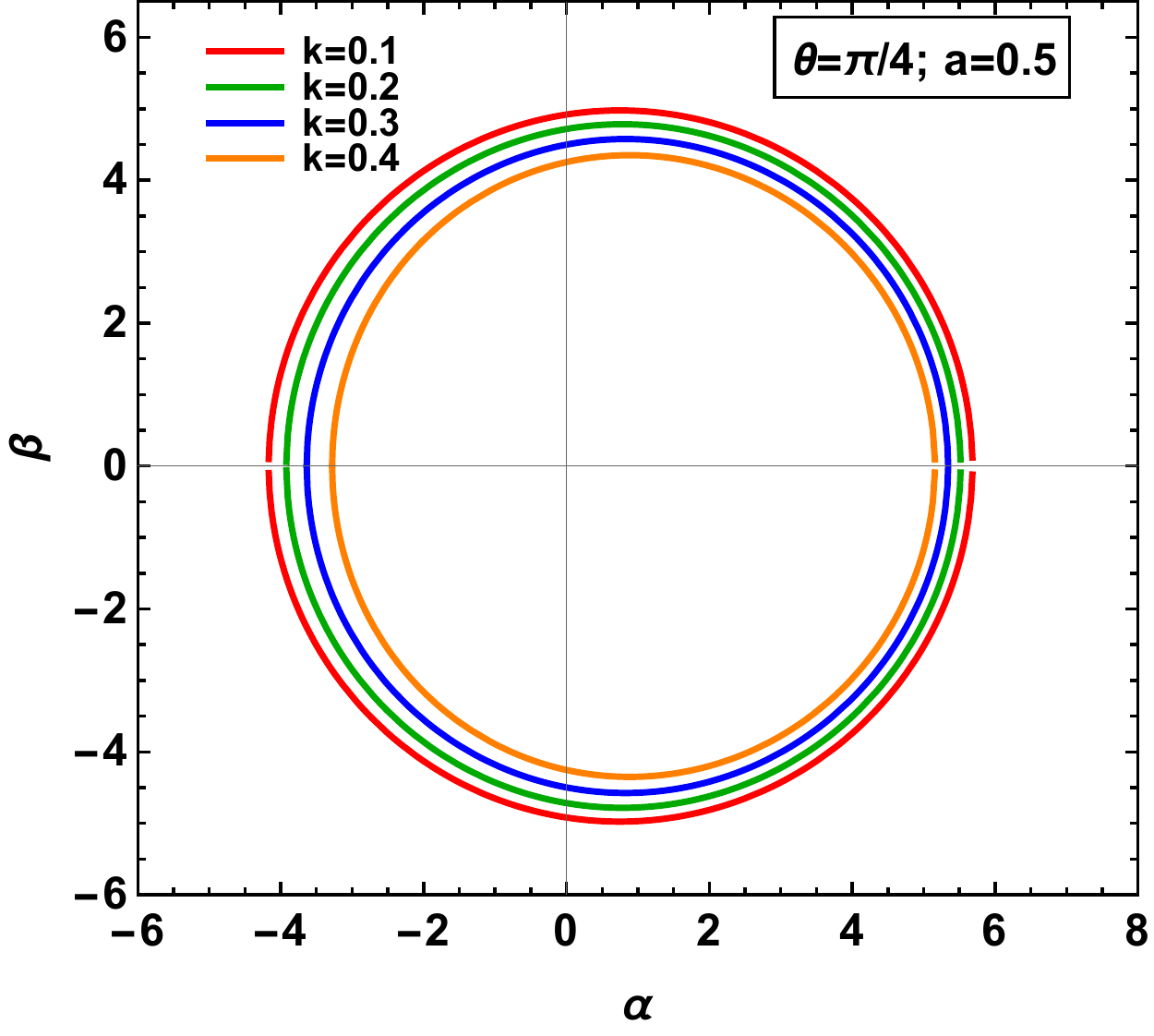} } \label{SP3_Fig_Vk}} 
   \qquad
     \subfloat[Variation of BH shadow with spin-parameter $a$. Here the inclination angle is taken to be $\theta=45^\circ$ and $k=0.1$ ]{{\includegraphics[width=7.5cm]{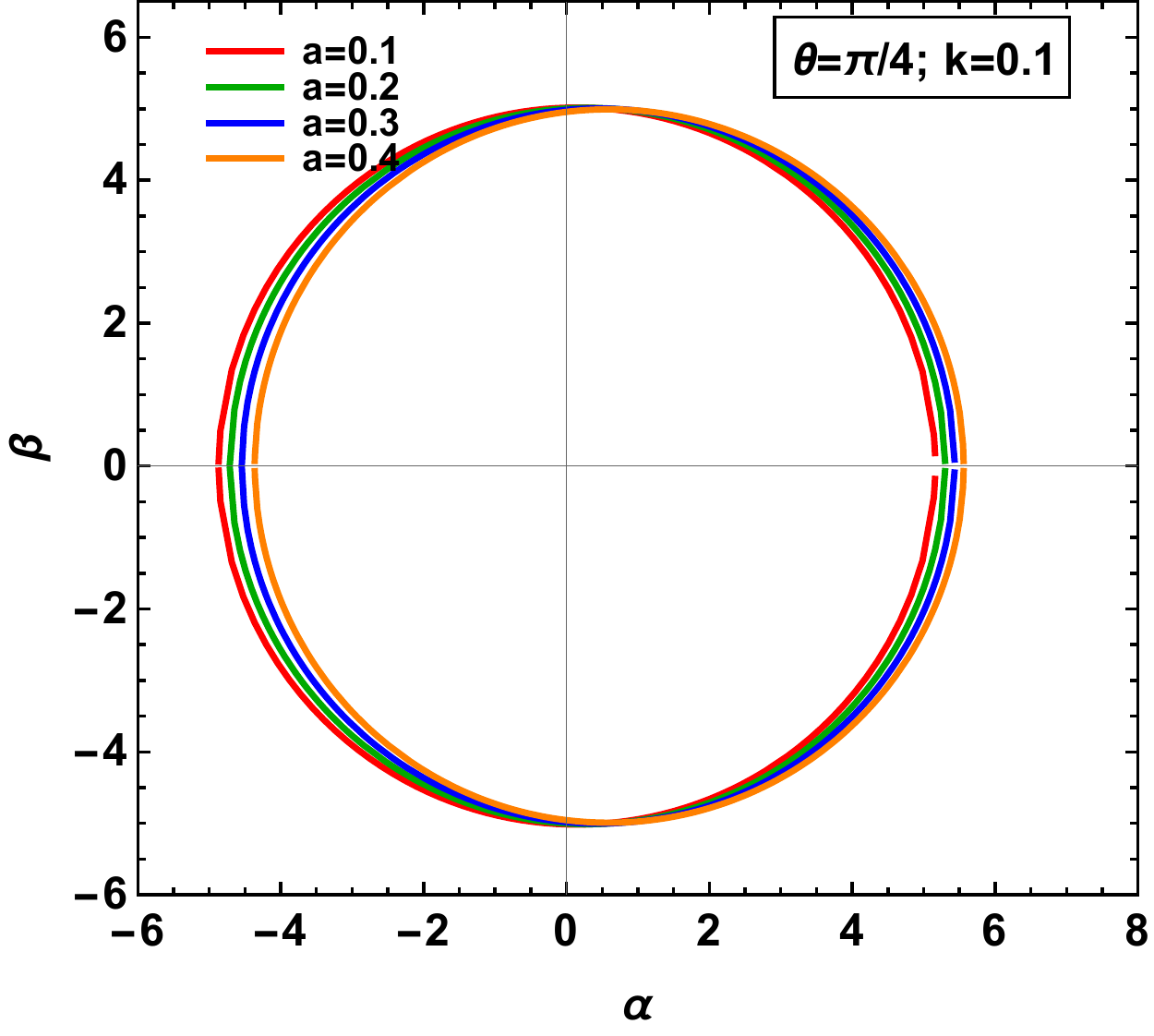} } \label{SP3_Fig_Va}}
   \qquad
    \subfloat[Variation of BH shadow with inclination angle $\theta$. Here the spin is taken to be $a=0.5$ and $k=0.1$]{{\includegraphics[width=7.5cm]{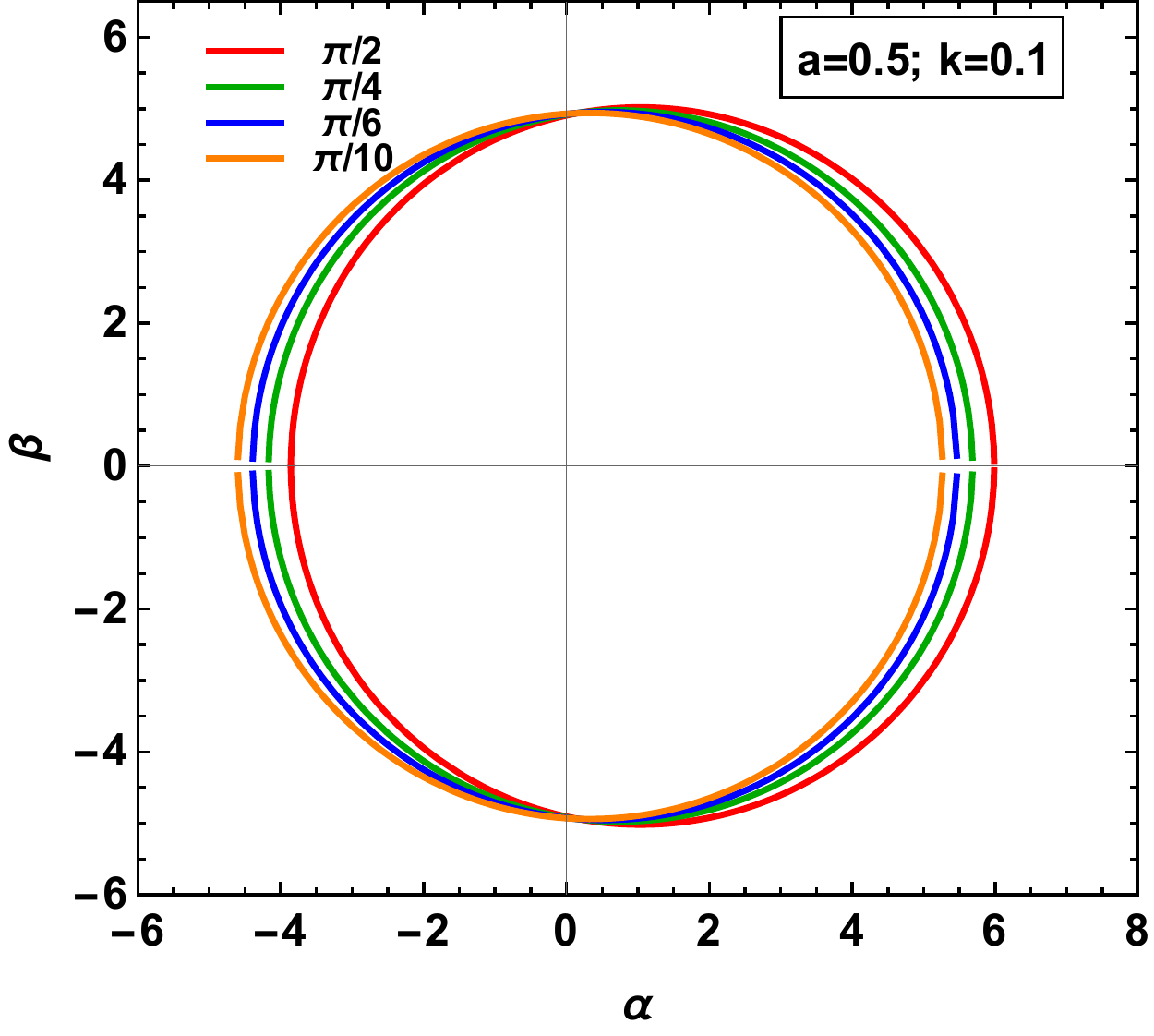} } \label{SP3_Fig_V4}}
   \caption{Variation of the shape and size of the BH shadow with charge parameter $k$, spin parameter $a$ and inclination angle $\theta$. We note that with increase in $\theta$ and $a$ the shadow becomes increasingly more dented.}
    \label{SP3_Fig_V}
\end{figure}

To derive this we need to consider the Bardeen tetrads \cite{Bardeen:1973tla,Carter:1968rr,1973ApJ...183..237C,2000CQGra..17..123D} which are associated with observers to whom the black hole appears non-rotating. These tetrads are given by,
\begin{align}
\label{24}
e_{(t)}&=\sqrt{g^{tt}}\partial_t + \frac{g^{t\phi}}{\sqrt{g^{tt}}}\partial_\phi \\
e_{(r)}&=\sqrt{|g^{rr}|}\partial_r \\
e_{(\theta)}&=\sqrt{|g^{\theta\theta}|}\partial_\theta \\
e_{(\phi)}&=\sqrt{|g^{\phi\phi}|+ \frac{(g^{t\phi})^2}{g^{tt}} }\partial_\phi
\end{align}
The components of four momentum $p_{(a)}$ of a locally inertial observer are $p_{(a)}=e_{(a)}^j p_j$ while $p^{(a)}=\eta^{(a)(b)}p_{(b)}$ such that,
\begin{align}
p^{(t)}&=\frac{E}{c}\Bigg(c\sqrt{g^{tt}}-\eta\frac{ g^{t\phi}}{\sqrt{g^{tt}}}\Bigg)\\
p^{(r)}&=\pm \sqrt{\frac{R}{\Sigma\Delta}} \\
p^{(\theta)}&=\pm \sqrt{\frac{\Theta}{\Sigma}} \\
p^{(\phi)}&=\sqrt{|g^{\phi\phi}|+ \frac{(g^{t\phi})^2}{g^{tt}} }\eta
\label{25}
\end{align}

Let $v_{(\theta)}=p^{(\theta)}/p^{(r)}$ and $v_{(\phi)}=p^{(\phi)}/p^{(r)}$ be the local apparent velocities of a given photon and $r_0$ and $\theta_0$ the Boyer-Lindquist coordinates of the observer. In that event the apparant perpendicular distance from the axis of rotation and the equatorial plane are respectively given by $d_\phi=r_0 v_{(\phi)}$ and $d_\theta=r_0 v_{(\theta)}$. Since $\dot{\theta}\to 0$ and $\dot{\phi}\to 0$ for $r\to \infty$ the $x$ and $y$ coordinates of the shadow are respectively given by,
\begin{align}
\alpha&=\lim_{ r_0\to \infty}\frac{r_0p^{(\phi)}(r_0,\theta_0)}{p^{(r)}(r_0,\theta_0)}=-\frac{\eta}{sin\theta_0} \nonumber \\
\beta&=\lim_{ r_0\to \infty}\frac{r_0p^{(\theta)}(r_0,\theta_0)}{p^{(r)}(r_0,\theta_0)}=\pm c \sqrt{\Theta(\theta_0)}
\label{26}
\end{align}
In \ref{SP3_Fig_V} we plot the variation of the shape and size of the black hole shadow with change in the inclination angle $\theta$, spin $a$ and for various choices of the non-linear electrodynamics charge parameter $k$. Similar study has been done in \cite{Tsukamoto:2014tja,Balart:2014cga,Amir:2016cen,Tsukamoto:2017fxq}. From the figure it is evident that an increase in $a$ and $\theta$ makes the shadow more deviated from the circular shape. Further, an increase in $k$ decreases the size of the shadow.

\section{Contact with observations}
\label{Sec4}
To find the effects beyond general relativity, we need to define observables associated with black hole shadow. 
\begin{figure}[t!]
\centering
\includegraphics[scale=0.3]{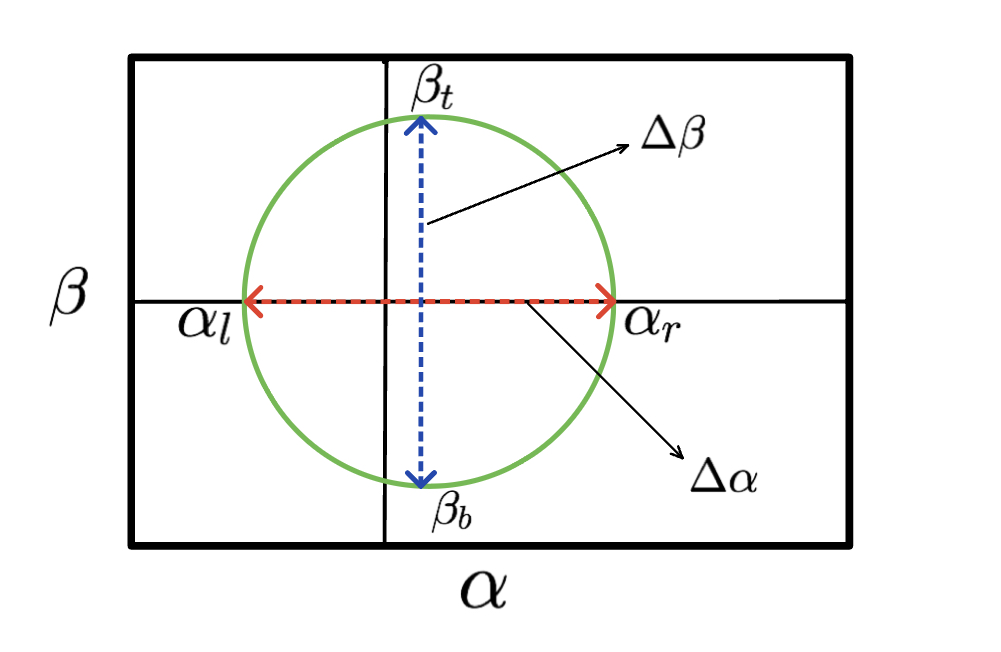}
\caption{Major axis and minor axis of shadow structure}
\label{Fig_IMG}
\end{figure}
The curve $\beta(\alpha)$ defines the boundary of the shadow. It is customary, for non-circular shadow structure, to define two axes (one major axis and one minor axis) corresponding to two diameters as shown in the \ref{Fig_IMG}. The major axis is given by $\Delta\beta=\beta_{t}-\beta_{b}$, where $\beta_{t}$ and $\beta_{b}$ are respectively the topmost and the bottommost points of the shadow in the $\alpha-\beta$ plane. In a similar way, the minor axis is defined by $\Delta \alpha=\alpha_{r}-\alpha_{l}$, where $\alpha_{r}$ and $\alpha_{l}$ are respectively the rightmost and the leftmost points of the shadow. One can note that the shadow structure is symmetric upon reflection about the minor axis. 
Further the two axes i.e $\Delta\beta$ and $\Delta\alpha$ can also be expressed as a function of metric parameters and the inclination angle. 
Therefore, observables used to distinguish the shadow of a black hole in modified gravity to that of the Kerr scenario in \gr\ are:\\
(i) Angular diameter which is given by:
\begin{align}
\Phi=\frac{GM \Delta	\beta}{c^2 D}
\label{Eq4-1}
\end{align}
where $M$ is the mass and $D$ is the distance from the observer.\\
(ii) The axis ratio which is denoted by:
\begin{align}
\Delta A=\frac{\Delta \beta}{\Delta \alpha}
\label{Eq4-2}
\end{align}
(iii) Deviation from circularity $\Delta C$: Unlike shadow of spherically symmetric metric, the shadow structure for Kerr-like solution looks like a deformed circle and hence it is important to define the observable, deviation from circularity. For this we first define the geometric centre of the shadow which can be evaluated as $\alpha_{c}=\int \alpha dS/ \int dS$ and $\beta_{c}=0$ (due to reflection symmetry around minor axis), where $dS$ is the area element. The distance between the geometrical centre and any point $(\alpha,\beta)$ on the shadow curve with azimuthal angle $\phi$ can be given by $\ell(\phi)=\sqrt{\left\{\alpha(\phi)-\alpha_{c} \right\}^{2}+\beta^{2}(\phi)}$. Now one can define the deviation from circularity as follows
\begin{flalign}
\Delta C= \dfrac{1}{R_{\rm avg}}\sqrt{\dfrac{1}{2\pi}\int_{0}^{2\pi}d\phi\left\{ \ell(\phi)-R_{\rm avg}\right\}}
\label{Eq4-3}
\end{flalign}
where, the average radius $R_{\rm avg}$ is defined as,
\begin{flalign}
R_{\rm avg}=\sqrt{\dfrac{1}{2\pi}\int_{0}^{2\pi}d\phi\, \ell^{2}(\phi)}
\label{Eq4-4}
\end{flalign}

\subsection{Constrains on the non-linear electrodynamics charge from the shadow of M87*}
\label{Sec4pt1}
In this section we compare our theoretical findings with the observed shadow of M87*. As reported by the Event Horizon Telescope (EHT) collaboration, the supermassive black hole M87* at the centre of the galaxy M87 has the following observational parameters \cite{Akiyama:2019cqa,Akiyama:2019fyp,Akiyama:2019eap}:

\begin{enumerate}
\item Angular Diameter: $42 \pm 3 \mu as$ \cite{Akiyama:2019cqa,Akiyama:2019fyp,Akiyama:2019eap}.
\item Deviation from circularity: $\Delta C\lesssim 10\%$ \cite{Akiyama:2019cqa,Akiyama:2019fyp,Akiyama:2019eap}.
\item Axis ratio: $\lesssim 4/3$ \cite{Akiyama:2019cqa,Akiyama:2019fyp,Akiyama:2019eap}.
\end{enumerate}
The mass of the source has been measured previously by investigating the motion of stars and gas clouds moving very close to the black hole. From stellar dynamics measurement the mass of the source is $M\simeq 6.2^{+1.1}_{-0.5}\times 10^9 M_\odot$ \cite{Gebhardt:2011yw}, while study of gas dynamics reveals that the mass of M87* is $M\simeq 3.5^{+0.9}_{-0.3}\times 10^9 M_\odot$ \cite{Walsh:2013uua}. The distance of the source has been estimated from stellar population measurements which turns out to be $D=(16.8\pm0.8)~\textrm{Mpc}$ \cite{Blakeslee:2009tc,Bird:2010rd,Cantiello:2018ffy}. Based on the angle the jet axis makes to the line of sight and assuming that the axis of rotation coincides with the jet axis, the inclination angle is estimated to be $17^{\circ}$ \cite{Akiyama:2019cqa,Akiyama:2019fyp,Akiyama:2019eap}. Further, from the measurement of the angular diameter of the shadow, the EHT collaboration reports the mass of the source to be $M=(6.5\pm 0.7)\times 10^{9}~M_{\odot}$.

\begin{figure}[H]
\begin{center}
\qquad
\subfloat[\label{Fig2a} Contours illustrating the dependence of the angular diameter of the shadow of M87* on $k$ and $a$ assuming $M\simeq 6.2\times 10^9 M_\odot$ and distance $D\simeq16.8~\textrm{Mpc}$]{\includegraphics[scale=0.65]{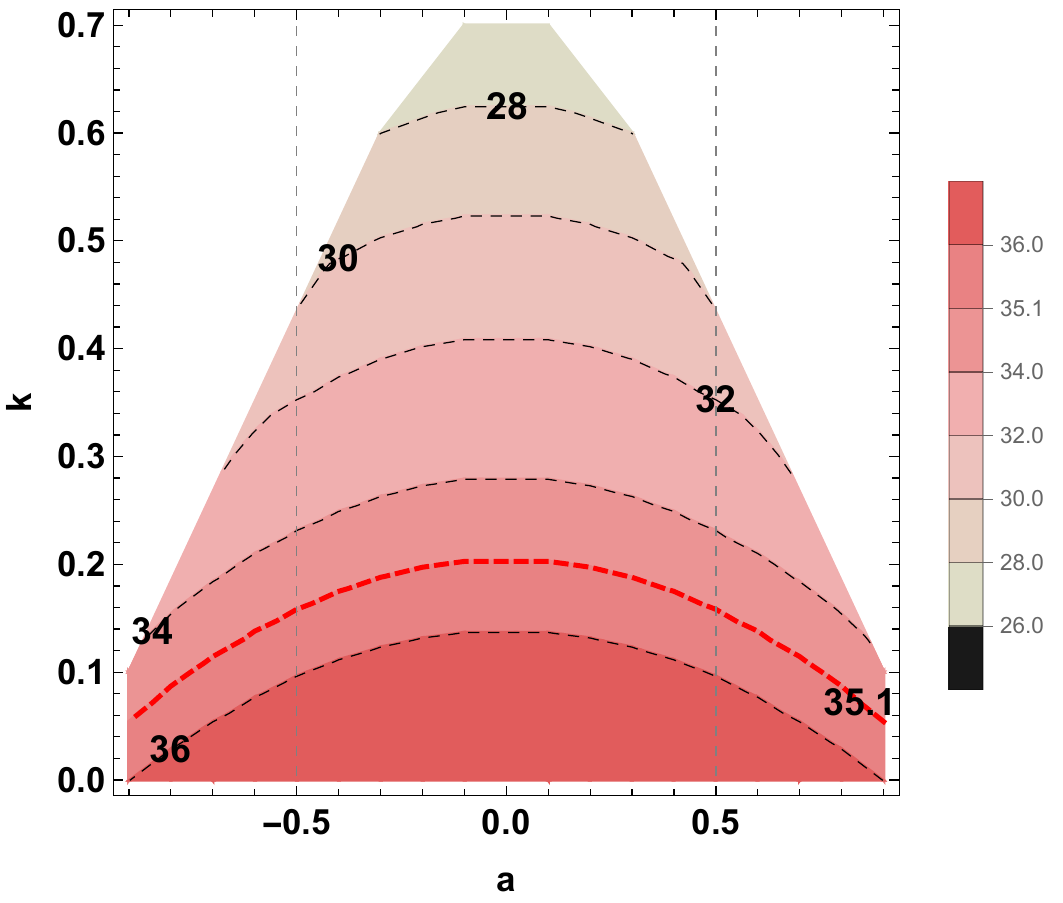}}
\qquad
\hspace{0.2cm}
\subfloat[Contours illustrating the dependence of the angular diameter of the shadow of M87* on $k$ and $a$ assuming $M\simeq 3.5\times 10^9 M_\odot$ and distance $D\simeq16.8\textrm{Mpc}$]{\includegraphics[scale=0.65]{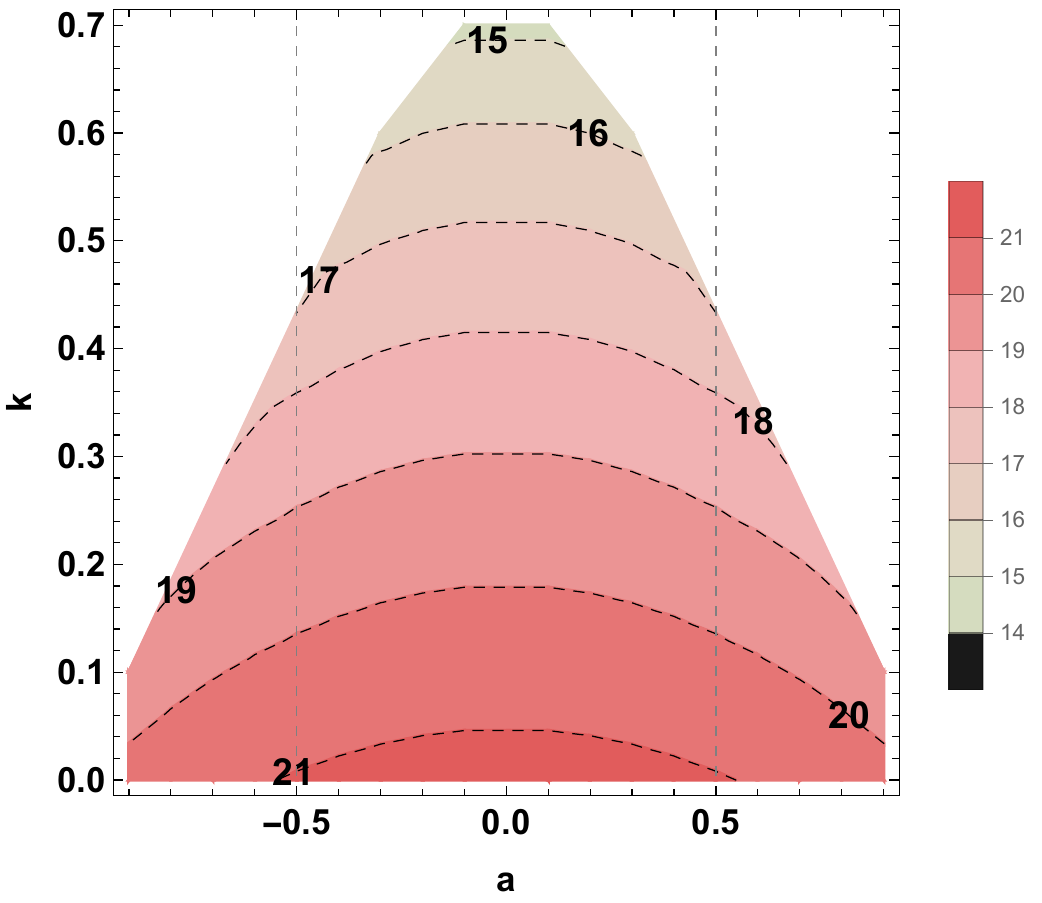}\label{Fig2b}
}

\caption{The above figure depicts the angular diameter of the shadow of M87* in the $(k,a)$ plane assuming the previously determined mass and distance.}
\label{Fig2}
\end{center}
\end{figure}

\begin{figure}[htb]
\begin{center}
\includegraphics[scale=0.6]{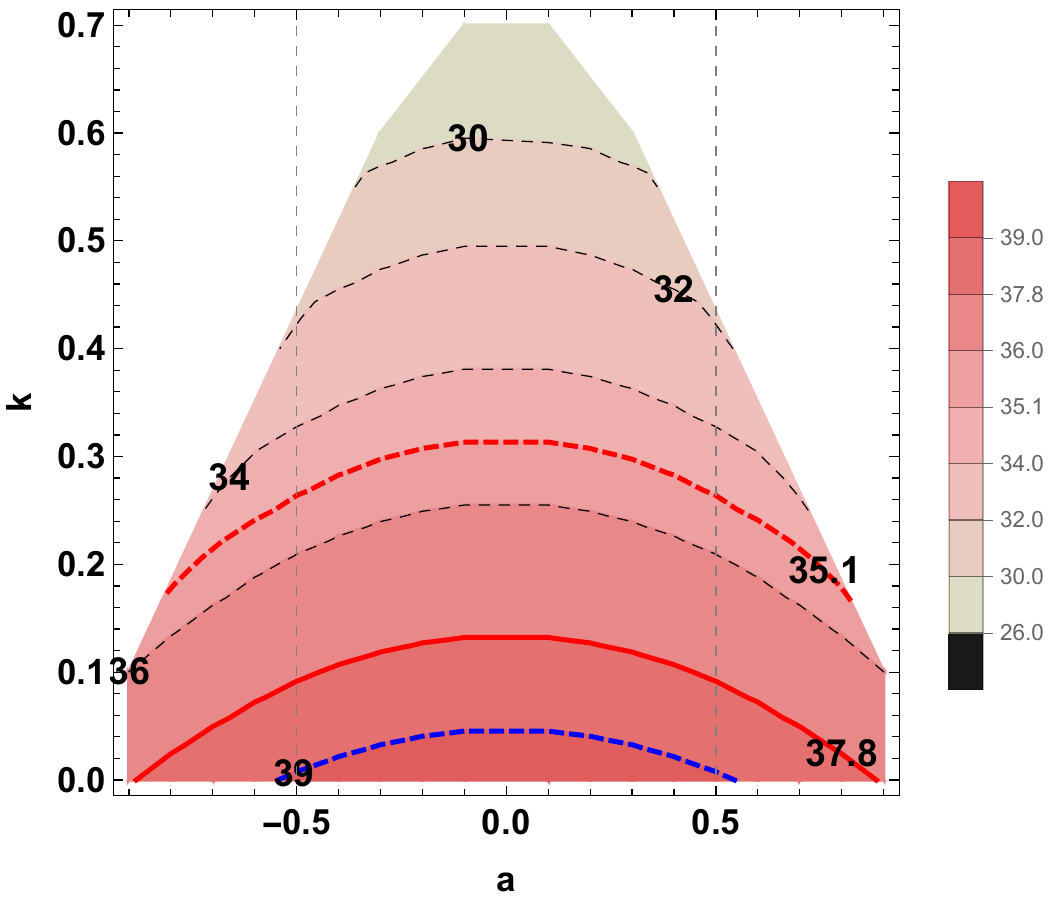}
\caption{In the figure above we plot the angular diameter of the shadow of M87* in the $(k,a)$ plane assuming distance $D\simeq16.8\textrm{Mpc}$ and mass $M\simeq 6.5\times 10^9 M_\odot$ (which is the mass estimated from the shadow itself).
}
\label{Fig3}
\end{center}
\end{figure}

It is important to note that in order to evaluate the angular diameter theoretically one needs to provide the mass and the distance (\ref{Eq4-1}) while the dependence on the background metric is encapsulated in $\Delta \beta$. 
In \ref{Fig2} and \ref{Fig3} we plot contours of angular diameter of M87* as a function of $a$ and $k$.   
In both the figures the angular diameter is evaluated based on the previously measured distance $D=16.8~\textrm{Mpc}$. In \ref{Fig2a} the angular diameter is calculated assuming $M\sim 6.2\times 10^9 M_\odot$ while the angular diameter in \ref{Fig2b} assumes $M\sim 3.5\times 10^9 M_\odot$. For completeness we report the angular diameter as function of $a$ and $k$ assuming $M=6.5\times 10^{9}~M_{\odot}$ in \ref{Fig3}  (which is the mass derived from the shadow itself). 

At this point it is important to recall that the measured angular diameter of the shadow is $42\pm 3 \mu as$. However, there is a 10\% offset between the image diameter and the shadow diameter, i.e the true shadow can be at most as small as $37.8\pm 2.7 \mu as$. Therefore, it is clear from \ref{Fig2b} that $M\simeq 3.5\times 10^9 M_\odot$ cannot reproduce the observed angular diameter of the shadow. It is clear from \ref{Eq4-1} that angular diameter is directly proportional to the black hole mass. Thus one might expect that a higher value of mass should reproduce the observed angular diameter. However, it is clear from \ref{Fig2a} that when $M\sim 6.2\times 10^9 M_\odot$ is taken, even the central value corresponding to the maximum 10\% offset $(37.8\mu as)$ cannot be reproduced. This might probably be the reason why the mass of M87* reported by the EHT Collaboration is greater than the previous estimates. We will discuss more on this towards the end of this section. 
From the rejection table of \cite{Akiyama:2019fyp} the spin values that can best explain the observations correspond to $|a|=0.5$ and $|a|=0.94$. Thus when $M\sim 6.2\times 10^9 M_\odot$ is used, a small value of $k$ ($0.05\lesssim k \lesssim 0.15$) barely reproduces the angular diameter of the shadow within $1-\sigma$ confidence interval, if we assume the maximum 10\% offset $(37.8-2.7)\mu as=35.1\mu as$. This is depicted by the red dashed line in \ref{Fig2a}. From \ref{SP3_Fig_V} it is evident that an increase in $k$ shrinks the shadow diameter and therefore the above discussion elucidates that the Kerr scenario or a small value of $k$ can better explain the observed angular diameter of M87*.

\begin{figure}[t]
\begin{center}
\subfloat[\label{Fig7a}The above figure shows the variation of $\Delta C$ with $k$ and $a$ assuming an 
inclination angle of $17^{\circ}$ corresponding to M87*.]
{\includegraphics[scale=0.65]{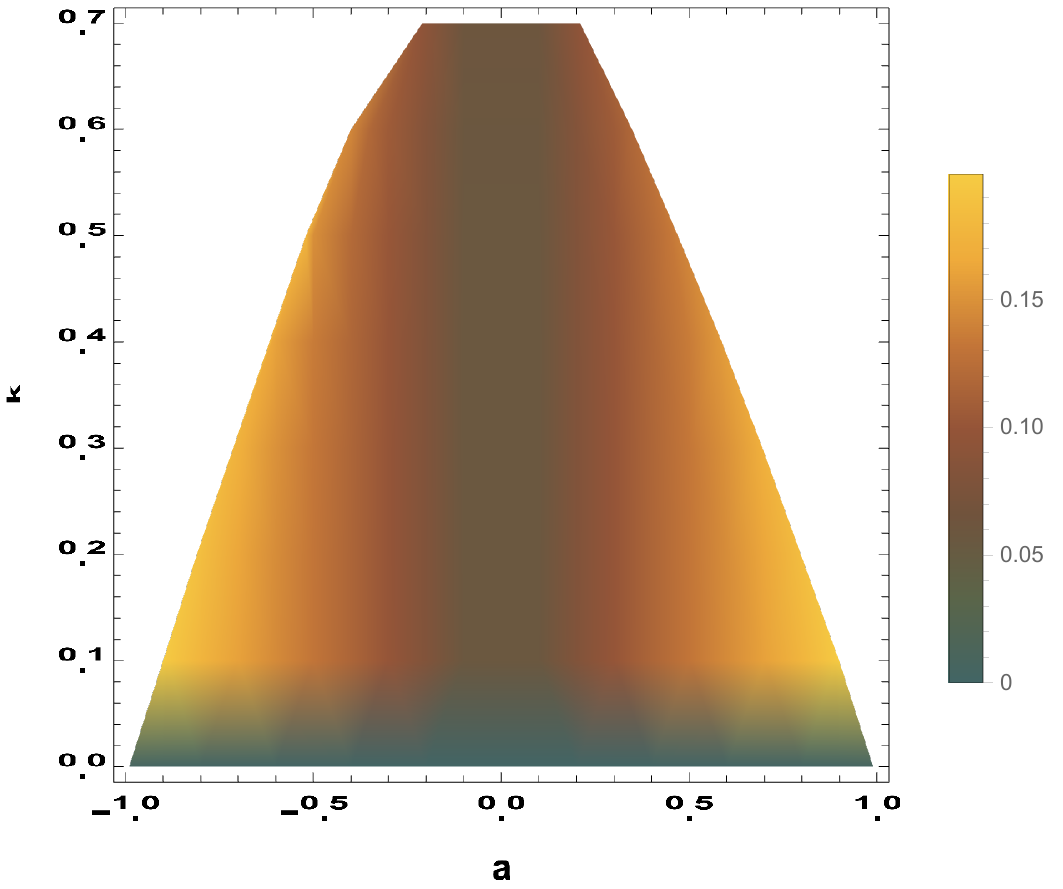}}
\hspace{0.3cm}
\subfloat[\label{Fig7b}The above figure shows the variation of $\Delta C$ with $k$ and $a$ assuming an inclination angle of $17^{\circ}$ corresponding to M87*.]
{\includegraphics[scale=0.65]{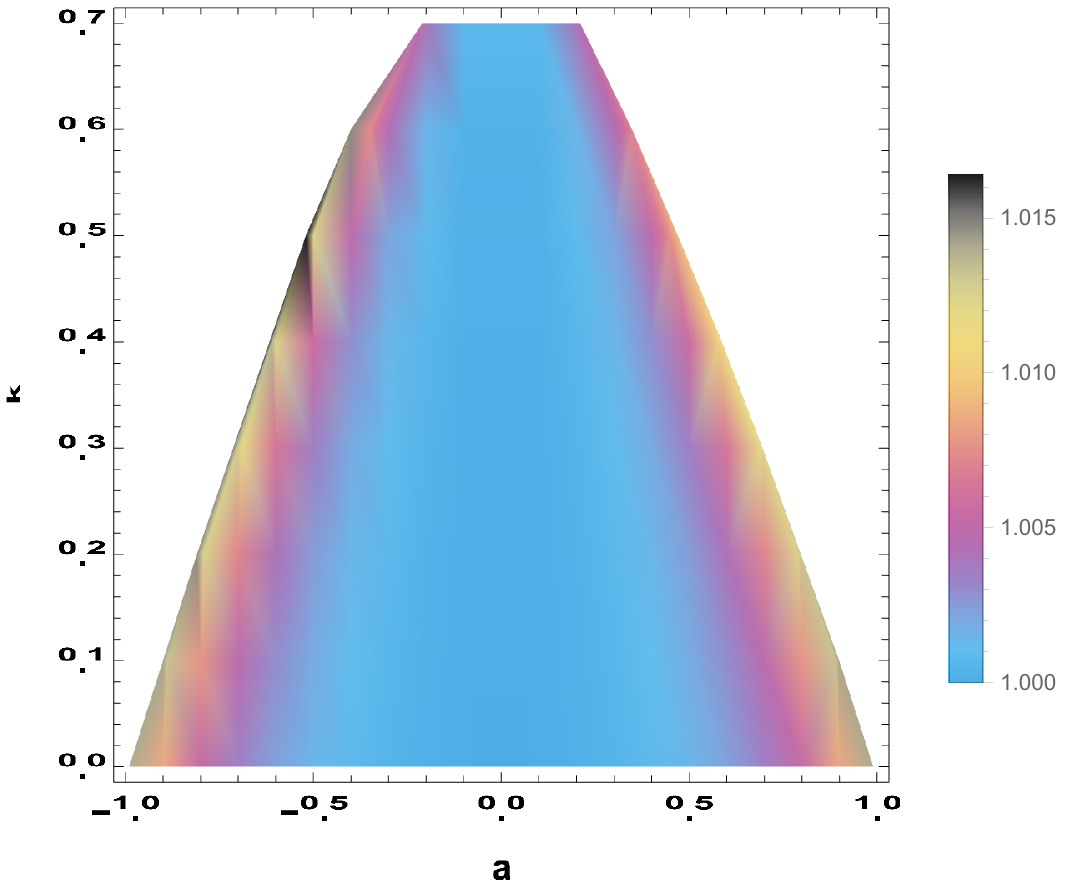}}
\caption{In the figure above we plot the variation of $\Delta C$ and $\Delta A$ as function of $k$ and $a$. The inclination angle is taken to be $17^\circ$ to obtain the plots.
}\label{Fig_DeltaC_M87}
\label{Fig4}
\end{center}
\end{figure}

Since an increase in mass increases the theoretical angular diameter of the shadow (\ref{Eq4-1}), when $M\simeq 6.5\times 10^9 M_\odot$ is assumed and $|a|>0.5$ is taken, $0\lesssim k \lesssim 0.1$ can reproduce the centroid value of the shadow angular diameter with 10\% offset. This is marked with red solid line in \ref{Fig3}. With this mass even a very small $k$ can reproduce the image diameter within 1-$\sigma$ ($39 \mu as$, marked with blue dashed line in \ref{Fig3}). However, $M\simeq 6.5\times 10^9 M_\odot$ should not be used to deduce the prefered value of $k$ from shadow related observations as this mass is itself derived from the shadow diameter.

It is important to note that the data related to deviation from circularity $\Delta C$ can establish independent constrains on the magnitude of $k$ (\ref{Fig7a}). Since $\Delta C \lesssim 0.1$, for low spin values, nearly all allowed values of $k$ seem to be favored by observations. However, the spin values that can best explain the observed jet power correspond to $|a|=0.5$ and $|a|=0.94$ (from the rejection table of \cite{Akiyama:2019fyp}). For $|a|\gtrsim 0.5$ the values of $k$ that satisfy $\Delta C \lesssim 0.1$ constrain correspond to $0\lesssim k\lesssim 0.1$ which is consistent with our conclusion from the observation related to angular diameter. Finally from \ref{Fig7b} we note that the third observable, i.e. axis ratio $\Delta \beta/\Delta \alpha\lesssim 4/3$ cannot provide any further constrain on the magnitude of $k$ \cite{Akiyama:2019cqa} as the maximum axis ratio realized in our case is 1.02.

\subsection{Constrains on the non-linear electrodynamics charge from the shadow of Sgr A*}\label{Sec4pt2}

\begin{figure}[]
\begin{center}

\subfloat[\label{Fig5a}The above figure depicts the angular diameter in the $k-a$ plane assuming $M=3.951\times 10^6 M_\odot$ and $D=7.935$ kpc.]{\includegraphics[scale=0.65]{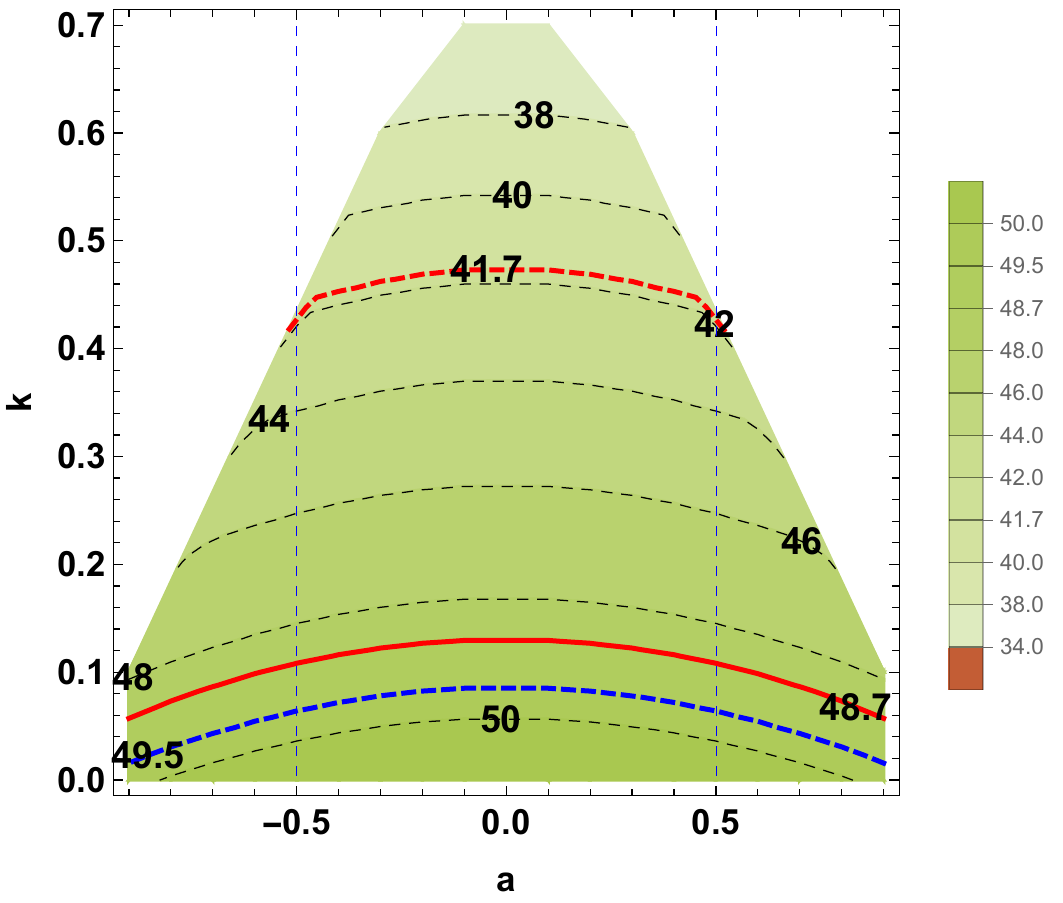}}
\hspace{0.36cm}
\subfloat[\label{Fig5b}The above figure depicts the angular diameter in the $k-a$ plane assuming $M=3.975\times 10^6 M_\odot$ and $D=7.959$ kpc.]
{\includegraphics[scale=0.65]{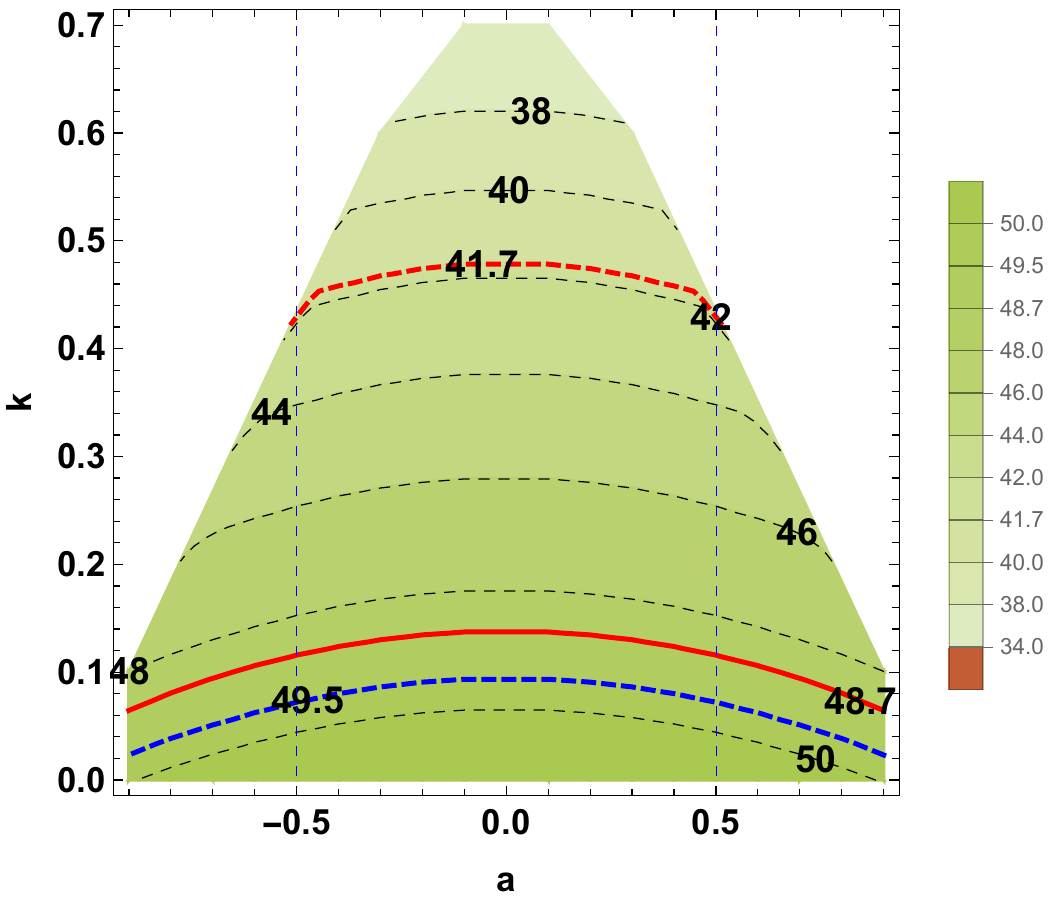}}
\\
\subfloat[\label{Fig5c}The above figure depicts the angular diameter in the $k-a$ plane assuming $M=4.261 \times 10^6 M_\odot$ and $D=8.2467$ kpc.]
{\includegraphics[scale=0.65]{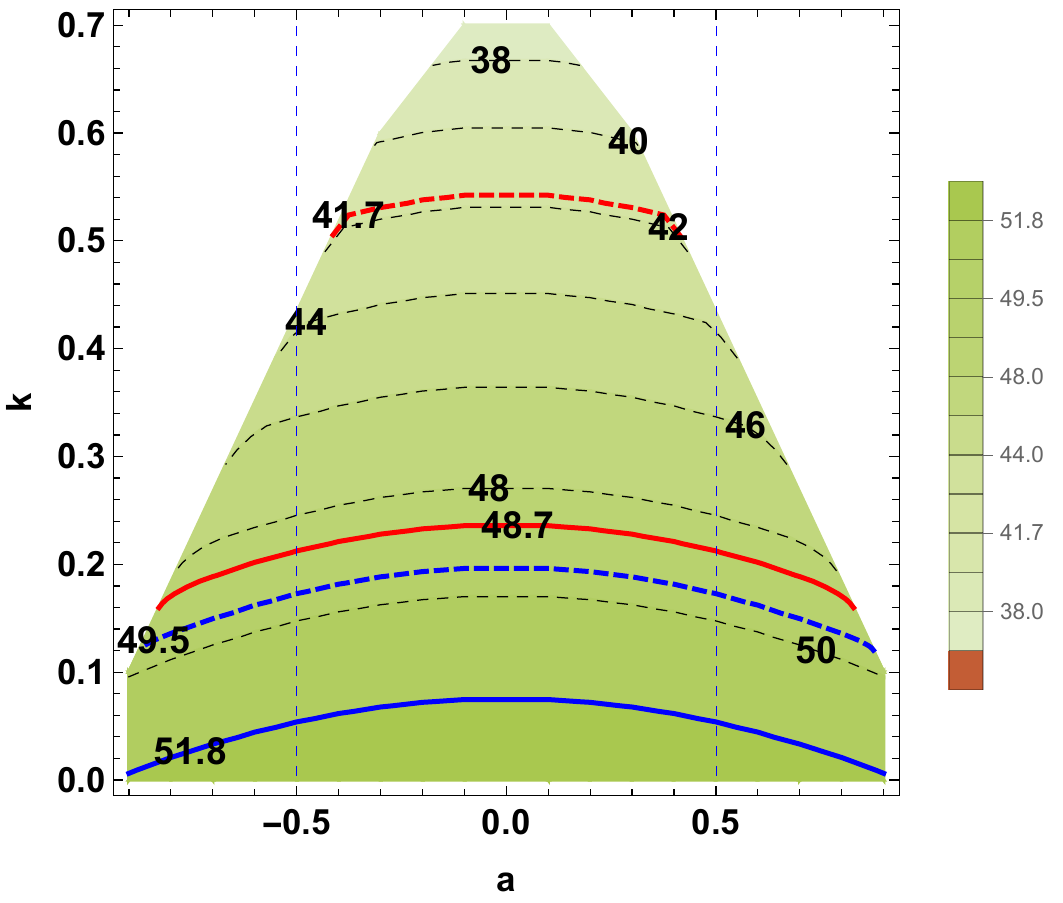}}
\hspace{0.36cm}
\subfloat[\label{Fig5d}The above figure depicts the angular diameter in the $k-a$ plane assuming $M=4.297\times 10^6 M_\odot$ and $D=8.277$ kpc.]
{\includegraphics[scale=0.65]{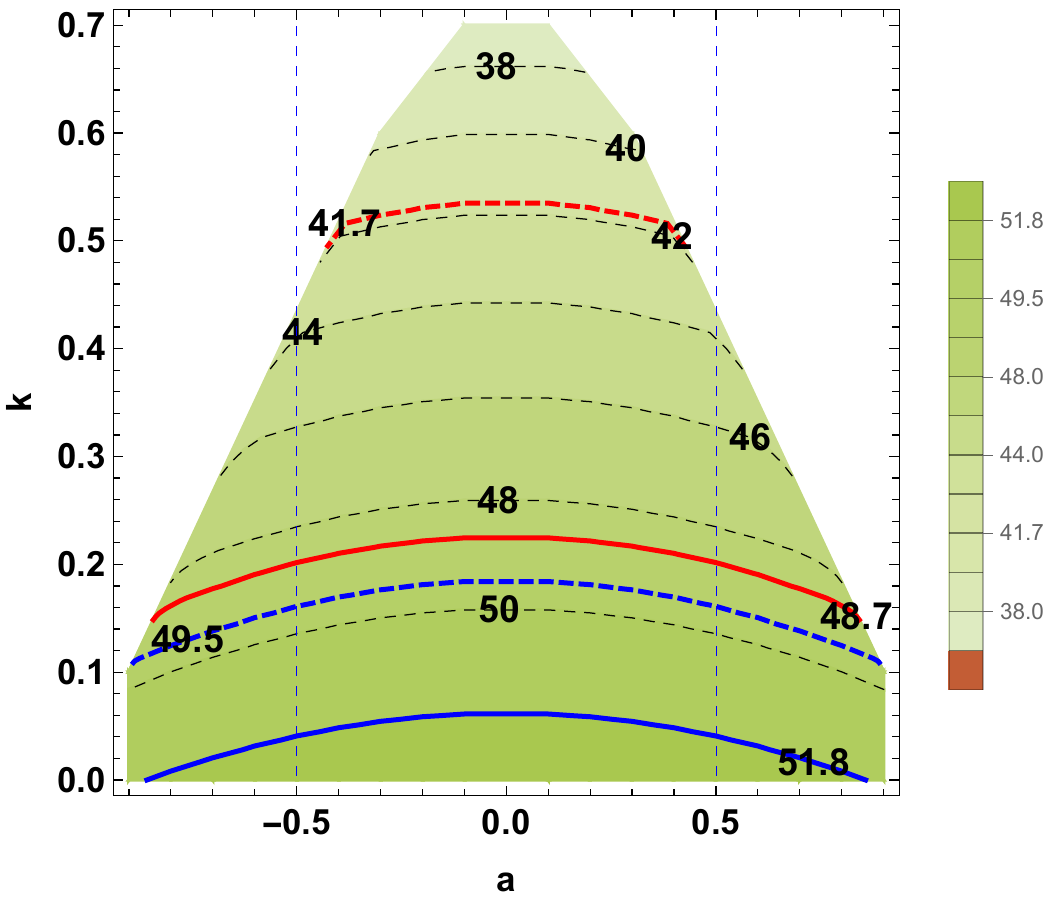}}

\caption{The above figure shows the variation of the angular diameter of Sgr A* as function of $k$ and $a$ using masses and distances reported by the Keck team and the Gravity collaboration. In all cases the inclination angle is taken to be $i=134^\circ$.}\label{Fig5}
\end{center}
\end{figure}

The recently observed shadow of Sgr A* can be used to establish constrains on the non-linear electrodynamics charge parameter, in the event the galactic centre black hole is considered to be a regular black hole with a Minkowski core. The mass and distance of the black hole is well constrained by several groups. Estimates on distance and mass measurements by the Keck team reports $D = 7959 \pm 59 \pm 32 $ pc and $M = (3.975 \pm 0.058 \pm 0.026) \times 10^6 M_\odot$ (for ﬁts that leave the redshift parameter free). Assuming the redshift parameter to be unity, the distance and mass measurements by the Keck team yield $D = 7935 \pm 50$ pc and $M = (3.951 \pm 0.047)\times 10^6 M_\odot$ respectively. The Gravity Collaboration estimates the distance and mass of Sgr A* to be $D=8246.7 \pm 9.3$ pc and $M = (4.261 \pm 0.012) \times 10^6 M_\odot$ respectively. By taking into account the systematics due to optical aberrations the GRAVITY collaboration constrains the mass and distance of Sgr A* to $M=4.297 \pm 0.012 \pm 0.040\times 10^6 M_\odot$ and $D=8277 \pm 9 \pm 33$ pc respectively.

The Event Horizon Telescope reports that the angular diameter of the emission ring of Sgr A* is $51.8\pm 2.3 \mu as$ while the shadow angular diameter is $48.7\pm 7 \mu as$.
By comparing the observed image of Sgr A* with models based on extensive numerical simulations, one concludes that the inclination angle of the source is $i<50^\circ$. In order to obtain the theoretical angular diameter of Sgr A* we need to use a specific value of the inclination angle. In this work we consider $i\simeq 134^\circ$ (or equivalently $46^\circ$) \cite{refId0}. We keep $k$ and $a$ variable and derive the angular diameter as a function of $k$ and $a$ using the aforesaid inclination angle and the various combinations of mass and distance discussed above. This is illustrated in \ref{Fig5}.

\begin{figure}[t]
\begin{center}
\subfloat[The above figure shows the variation of $\Delta C$ with $k$ and $a$ assuming an 
inclination angle of $134^{\circ}$ corresponding to Sgr A*.]
{\includegraphics[scale=0.65]{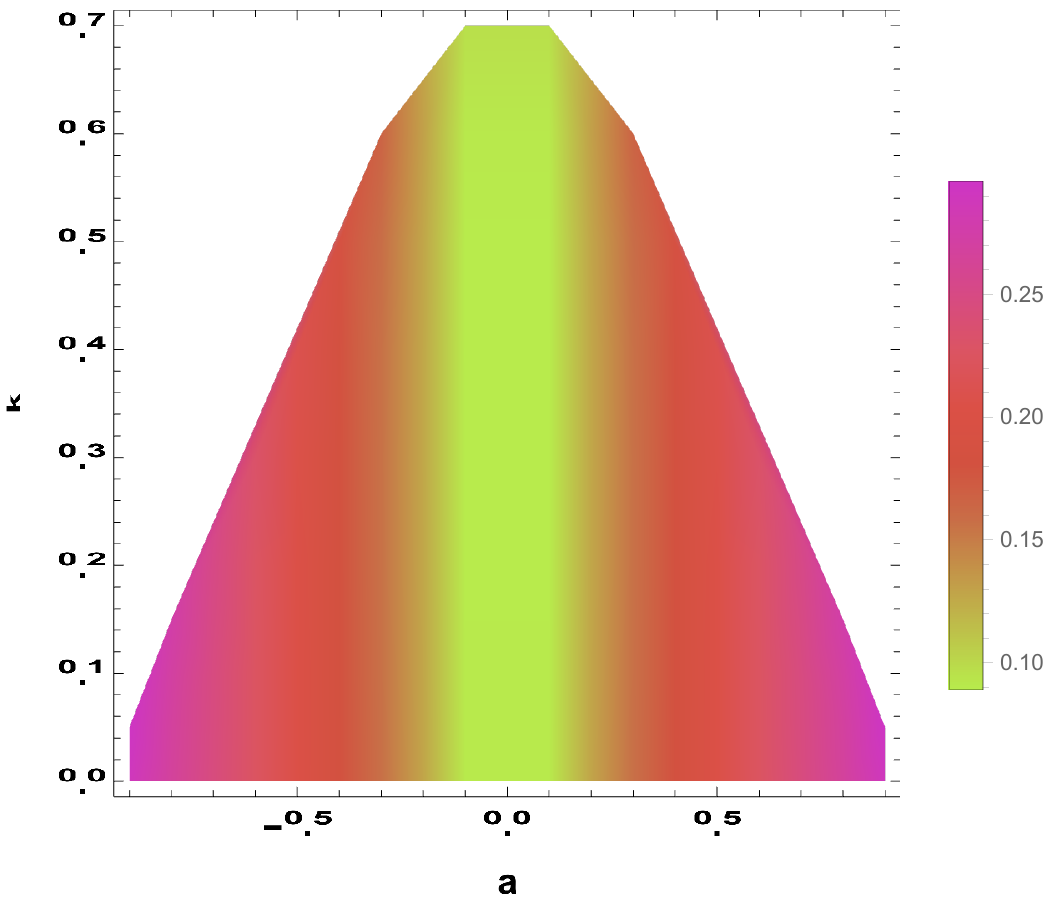}}
\hspace{0.3cm}
\subfloat[The above figure shows the variation of $\Delta A$ with $k$ and $a$ assuming an inclination angle of $134^{\circ}$ corresponding to Sgr A*.]
{\includegraphics[scale=0.65]{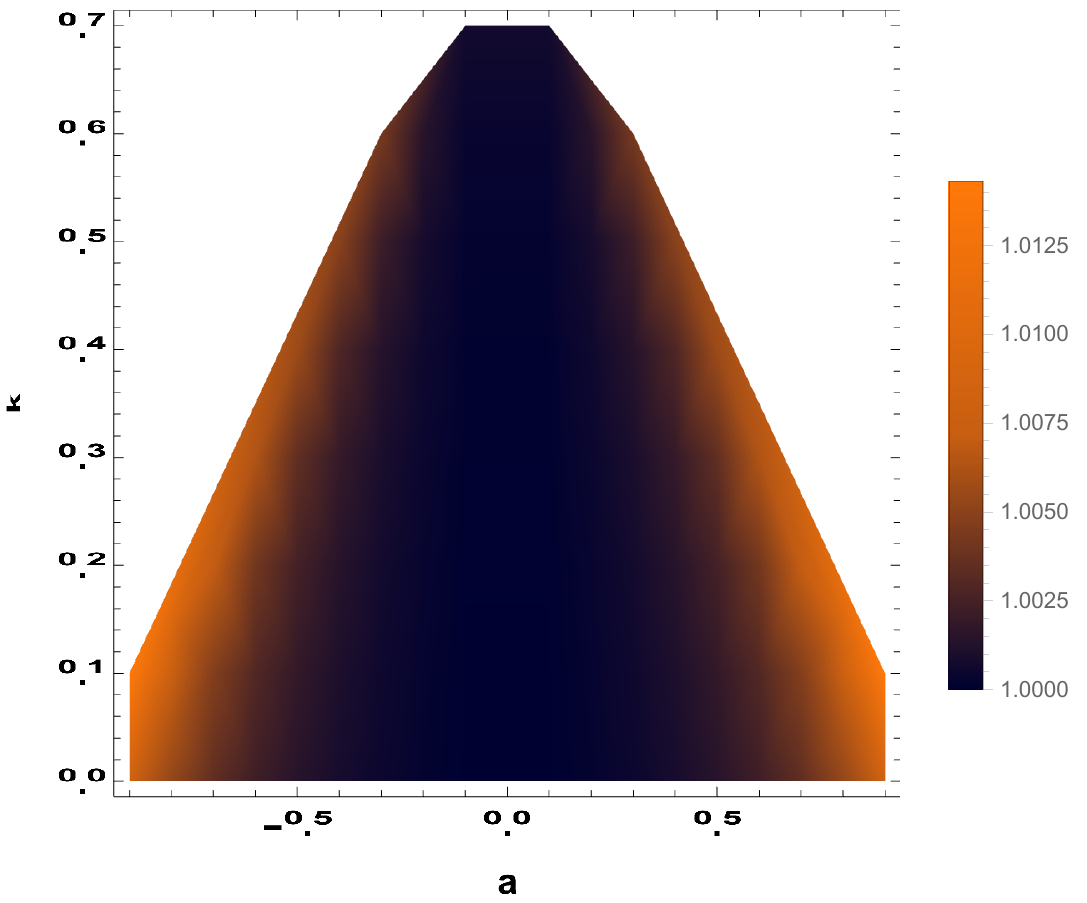}}
\caption{In the figure above we plot the variation of $\Delta C$ and $\Delta A$ as function of $k$ and $a$, for the source Sgr A*. The inclination angle is taken to be $134^\circ$ to obtain the plots.
}\label{Fig_DeltaC_M87}
\label{Fig6}
\end{center}
\end{figure}

We note from \ref{Fig5} that in order to reproduce the central value of the angular diameter of the shadow ($48.7 \mu as$), $k$ as high as $0.1$ is allowed (red solid line in \ref{Fig5a} and \ref{Fig5b}) if we consider the following combinations of mass and distance: (a) 
$M = (3.975 \pm 0.058 \pm 0.026) \times 10^6 M_\odot$ and $D = 7959 \pm 59 \pm 32 $ pc, (b)
$M = (3.951 \pm 0.047)\times 10^6 M_\odot$ and $D = 7935 \pm 50$ pc. When the mass and distance reported by the Gravity collaboration is used, $k$ as high as $0.2$ is allowed (from the red solid line in \ref{Fig5c} and \ref{Fig5d}).
If we wish to reproduce the angular diameter of the shadow upto 1-$\sigma$ (red dashed line \ref{Fig5}) then all the four combinations of mass and distance allow $k$ as high as 0.5 (\ref{Fig5}). However, we do not emphasize much on this result as the error bar of $\pm 7$ associated with the shadow diameter is quite high. Therefore, when shadow diameter is considered a small but non-trivial value of $k$ explains the observation better than the Kerr scenario. 

However, when the image diameter is considered $k$ as high as 0.05 is allowed within 1-$\sigma$ (blue dashed line in \ref{Fig5a} and \ref{Fig5b}) if we assume mass and distance measurements by the Keck collaboration. Therefore, the Kerr scenario can reproduce the image diameter better in this case.
When the mass and distance reported by the Gravity collaboration is used to calculate the angular diameter, then a small but non-trivial $k\simeq 0.05$ can reproduce the central value of $51.8\mu as$ (blue solid line in \ref{Fig5c} and \ref{Fig5d}) while $k\simeq 0.15$ is allowed within 1-$\sigma$ (blue dashed line in \ref{Fig5c} and \ref{Fig5d}). Hence, in this case the regular black hole scenario is more favored compared to the Kerr scenario.

From the above discussion we note that the present observation related to the shadow of Sgr A* generally favors a small value of the non-linear electrodynamics charge prameter $k$.  
We also plot the variation of the axis ratio $\Delta A$ and the deviation from circularity $\Delta C$ for the shadow of Sgr A* as function of $k$ and $a$. This is illustrated in \ref{Fig6}. At present we do not have constraints on $\Delta A$ and $\Delta C$ from the shadow of Sgr A*. In future when these are reported we can establish independent constraints on $k$ and $a$, just as we did for M87*.

\section{Concluding Remarks}\label{Sec6}
The present work investigates the role of the non-linear electrodynamics charge in explaining the recently observed shadow of M87* and Sgr A*. Such a charge arises in Einstein gravity coupled to non-linear electrodynamics and can potentially evade the $r=0$ curvature singularity, which inevitably arises in GR. Such regular black holes are endowed with a de Sitter or a Minkowski core. The theoretical implications and observational consequences of regular black holes with a de Sitter core have been extensively studied in the past. This motivates us to explore the observational signatures of regular black holes with a Minkowski core. Such black holes have an exponential mass function which makes the corresponding quantum gravity theory finite to all orders. 

We work out the nature of the black hole shadow for a general stationary and axisymmetric metric and subsequently specialize to the regular black hole scenario with a Minkowski core. 
Presence of an event horizon demands that the non-linear electrodynamics charge parameter $k$  varies in the range $0\lesssim k \lesssim 0.7$. 
Our study reveals that with increase in $k$ the shadow becomes smaller in size. 
Non-rotating black holes and black holes viewed at $0^\circ$ inclination angle cast a circular shadow. Deviation from circular shape occurs with increase in black hole spin or the inclination angle. This enables us to define a major axis ($\Delta\beta$) and a minor axis ($\Delta\alpha$) for the shadow which depends on the metric parameters $k$ and $a$.

Derivation of the theoretical outline of the shadow enables us to compute the shadow related observables, namely, the angular diameter, the deviation from circularity and the axis ratio which are subsequently compared with the observed shadow of M87* and Sgr A*. Computation of the angular diameter requires one to provide information regarding the mass and the distance, e.g. an increase in mass enhances the shadow angular diameter. 
Thus, it turns out that for M87* when $M=3.5 \times 10^9 M_\odot$ (obtained from gas dynamics measurements) is used to calculate the theoretical angular diameter, it cannot reproduce the observed angular diameter of $42\pm 3 \mu as$. Even when the mass of M87* based on stellar dynamics observation is used (i.e. $M=6.2\times 10^9 M_\odot$, which is considerably larger than the previous one) the observed angular diameter of the shadow cannot be reproduced. However, if one considers the maximum offset of 10\% in the shadow diameter then $0.05\lesssim k \lesssim 0.15$ can reproduce the observed angular diameter within 1-$\sigma$.
Therefore, the shadow of M87* can be explained by the regular black hole scenario with a  small but non-zero value of $k$.  
It is interesting to note that the shadow angular diameter scales with the mass of the black hole. This might be the plausible reason the mass of M87* derived from the observed angular diameter (of $42 \pm 3 \mu as$) is $6.5\times 10^9 M_\odot$ which is greater than both the previous estimates.

Mass and distance of Sgr A* have been well constrained by the Keck team and the Gravity collaboration. When mass and distance reported by the Keck team is used to compute the theoretical angular diameter, we note that $0.05\lesssim k \lesssim 0.1$ best explains the central value of the observed shadow angular diameter. Again when mass and distance reported by the Gravity collaboration is used to evaluate the theoretical angular diameter, $0.15\lesssim k \lesssim 0.2$ is required to address the observation of $48.7\mu as$. Therefore, for the case of Sgr A* the regular black hole scenario is more favored compared to the Kerr scenario. In particular, a small but non-trivial value of $k$ is required to reproduce the observed angular diameter of Sgr A*. 

It is interesting to note that the present results are consistent with our previous works where we established constrains on $k$ from observations associated with QPOs \cite{Banerjee:2022chn} and the continuum spectrum \cite{Banerjee:2022len}. From the QPO related observations we found that some QPO models favor the Kerr scenario while most of them indicate that a small value of $k$ explains the observations better. The optical observations of quasars on the other hand exhibit preference towards the Kerr scenario. It is important to note that none of the observations (shadow, QPO or continuum spectrum) favor a high value of $k$. This is interesting as these works are based on completely different observational samples and they collectively indicate towards the same finding. This outcome can be subjected to further test with the availability of more and more images of black holes with increasingly better resolution.




 
\section*{Acknowledgements}

The research of SSG is partially supported by the Science and Engineering Research Board-Extra Mural Research Grant No. (EMR/2017/001372), Government of India. Research of I.B. is funded by the Start- Up
Research Grant from SERB, DST, Government of India
(Reg. No. SRG/2021/000418).

\bibliography{Black_Hole_Shadow,Brane,KN-ED,regularBh,regularBh2,bardeen,SgrA,QPO,QPO2,IB}

\bibliographystyle{./utphys1}
\end{document}